\documentclass{article}

\usepackage[utf8]{inputenc}
\usepackage{natbib}
\usepackage{amsmath}
\usepackage{graphicx}
\usepackage{grffile}
\usepackage{booktabs}
\usepackage{subfigure}
\usepackage[boxed]{algorithm}
\usepackage{algorithmic}
\usepackage{multicol}
\usepackage{tikz}
\usepackage{mdwlist}
\usepackage{flushend}
\usepackage{psfrag}
\usepackage[hyphens]{url}
\usepackage[shortcuts]{extdash}
\usepackage{soul}

\newcommand{\experimentNetwork}{Pretty Good Privacy }
\newcommand{\experimentNetworkSize}{10{,}680 }
\newcommand{\experimentNetworkCount}{36 }
\newcommand{\experimentNetworkInternal}{arenas-pgp}
\newcommand{\experimentNetworkCite}{\cite{konect:boguna}}

\title{
  Guided Graph Generation:
  Evaluation of Graph Generators in Terms of Network Statistics, and a New Algorithm
}

\author{
  Jérôme Kunegis\textsuperscript{1,2,3}, Jun Sun\textsuperscript{2} \& Eiko Yoneki\textsuperscript{3} \\
  \texttt{jerome.kunegis@unamur.be, junsun@uni-koblenz.de,} \\ 
  \texttt{eiko.yoneki@cl.cam.ac.uk} \\
  \textsuperscript{1} University of Namur, Belgium \\
  \textsuperscript{2} University of Koblenz--Landau, Germany \\
  \textsuperscript{3} University of Cambridge, United Kingdom
}

\begin{document}

\maketitle

\begin{abstract}
  We consider the problem of graph generation guided by
  network statistics, i.e., the generation of graphs which have given
  values of various numerical measures that characterize networks,
  such as the clustering coefficient and the number of cycles of
  given lengths.   
  Algorithms for the generation of synthetic graphs are often based on
  graph growth models, i.e., rules of adding (and sometimes removing)
  nodes and edges to a graph that mimic the processes present in real-world
  networks.  While such graph generators are desirable from a
  theoretical point of view, they are often only able to reproduce a
  narrow set of properties of real-world networks, 
  resulting in graphs with otherwise unrealistic properties. 
  In this article, we instead evaluate common graph generation
  algorithms at the task of reproducing the numerical statistics of
  real-world networks, such as the clustering coefficient, the degree
  assortativity, and the connectivity. 
  We also propose an iterative algorithm, the Guided Graph Generator, 
  based on a greedy-like procedure that recovers realistic values over a
  large number of commonly used graph statistics, while at the same time
  allowing an efficient implementation based on incremental updating of
  only a small number of subgraph counts.  We show that the proposed algorithm
  outperforms previous graph generation algorithms in terms of the error
  in the reconstructed graphs for a large number of graph statistics
  such as the clustering coefficient, the assortativity, the mean node
  distance, and also evaluate the algorithm in terms of precision, speed
  of convergence and scalability, and compare it to previous graph
  generators and models.  We also show that the proposed algorithm generates
  graphs with realistic degree distributions, graph spectra, clustering
  coefficient distributions, and distance distributions.
\end{abstract}

\section{Introduction}
The problem of graph generation is concerned with finding algorithms
that generate graphs whose properties match those of real networks.  
In general, graph generators are made to be realistic in two ways:  (1)~by
mimicking the temporal evolution seen in a given input graph, and (2)~by
reproducing statistical properties of a given input graph.  The two
approaches are connected in a nontrivial fashion:  A realistic graph
growth model should in principle lead to realistic structural graph
properties.  In practice however, only the simplest models allow this
relationship to be derived in closed form, and a practical graph
generation algorithm can then usually follow only one of both.  Due to the
simplicity of generating graphs edge by edge and node for node, a large number of graph
generators are thus formulated to follow criterion~(1), without being
able to derive guarantees for criterion~(2).  As a result, most graph
generators cannot be easily tuned to produce given graph statistics.
For instance, while many graph generators have a parameter that controls
the amount of clustering (e.g., the probability of forming a triangle),
these parameters cannot be easily adjusted to result in a requested
value of the clustering coefficient -- making such algorithms unsuited
for generating graphs with an exact value of the clustering
coefficient.  This situation becomes even more difficult when multiple
numerical graph properties are considered simultaneously. 
As an example, to generate a graph with a given degree
distribution and clustering coefficient, a popular strategy involves
first generating a graph with the requested degree distribution (via
random assignment of half-edges), and then exchanging individual edges
(performing \emph{switches}) in a way that does not change the degree
distribution, but changes the clustering coefficient.  These techniques
can be extended to switches of more than four nodes, as done for instance by Bansal
and colleagues \citeyearpar{gp44}.\footnote{ This includes the
  \emph{Big-V} method as a special case \citep{gp55}.  }  These kinds of
methods are, by their nature, not generalizable to arbitrary graph
statistics, since individual switching moves are restricted to
maintaining a small set of graph properties.  
Therefore, this article will evaluate graph generation algorithms in
terms of their ability to generate graphs with given values of numerical
graph properties, and propose a new graph generation algorithm (the
Guided Graph Generator) designed
to achieve this goal with a high precision. 
The algorithm presented in this article is incremental and greedy-like, and uses the
principle that no graph statistic should be taken as fixed~--
as long as the graph as a whole becomes closer to its
intended statistic values.  We show that the Guided Graph Generator can
generate graphs that match the requested properties very closely,
outperforming previous algorithms by several orders of magnitude in
terms of precision.

Generating graphs with given properties is a central problem in the area
of complex network analysis and graph mining, and can be used for various
purposes: (1)~anonymizing a network: generating a network with similar
properties to a given one, but in which details of the original network
cannot be recovered, (2)~sampling a network: generating a network
smaller than a given network, but with otherwise similar properties;
this allows one to apply computationally expensive network analysis
methods to networks that would normally be too large, and (3)~scalability
testing: generating graphs larger than a given graph, for the purpose of
testing the scalability of a given network analysis method.  Thus, it is
no surprise that many different graph generators have been investigated,
and that an algorithm for generating graphs that match the properties of
real-world graphs well is an important goal.  
In all three cases, it is also true that generating graphs with precise
values of the statistics is more important than having realistically
designed steps of the algorithm, since in most cases only the final
output of these algorithms is used. 
In this paper, we therefore propose
a graph generation algorithm based on the idea of matching the value of
numerical statistics: The graphs it generates have the same statistic
values as any given input graph.  This means that the proposed algorithm has the
desirable property of having a very small parameter space.
Nevertheless, it is able to reproduce given statistic values to a
very high precision, so high in fact that, as we will see, other properties of the
graph are also reproduced faithfully.  Another concern with graph models is
tractability:  While graph models based on arbitrary numerical
statistics exist (i.e., exponential random graph models, which represent
one generic solution to the problem presented in this paper), the resulting
fitting and graph generation algorithms are non-scalable to the point of
being unused in practice for large graphs.  The algorithm presented here has no such
limitation, as an efficient method for vectorizing its
calculations is available.

We start the paper by reviewing graph statistics and
graph models in Section~\ref{sec:background}, with a focus on deriving
the graph statistics resulting from each model.  Then,
in Section~\ref{sec:algorithm}, we state the proposed Guided Graph
Generator algorithm.  We  
evaluate the previously existing and the proposed algorithm in
Section~\ref{sec:experiments} in terms of 
precision, scalability, and ability to reproduce various characteristic graph
distributions, as well as compare the convergence of the proposed
algorithm to the behavior of the previously existing algorithms. 
In Section~\ref{sec:limitations}, we discuss the limitations of the
approach, and conclude in Section \ref{sec:conclusion}. 

\section{Graph Generation}
\label{sec:background}
A synthetic graph, as opposed to a real-world graph, is a graph
generated algorithmically.  
A survey of graph generation algorithms is given by Chakrabarti and Faloutsos \citeyearpar{b876}.
Graph generation algorithms can have many orthogonal goals, not all of
which are shared by our approach -- in the following, we give a
structured overview of such graph generation 
algorithms with a focus on how each relates to the statistics of the
graphs it generates.

\subsection{Graph Statistics}
As used in this article, a network statistic is a numerical value
that characterizes a network.  Examples of network statistics are the
number of nodes and the number of edges in a network, but also more
complex measures such as the diameter and the clustering coefficient.
Statistics are the basis of a very large class of network analysis methods; they can be
used to compare networks, classify networks, detect anomalies in
networks and for many other tasks.  Network statistics are also used
to map a network's structure to a simple numerical space, in which
many standard statistical methods can be applied.  Thus, network
statistics are essential for the analysis of almost all network types.
All statistics considered in this article are real numbers. 

By definition, any graph model reduces a graph to certain
characteristics of a graph.  These can be individual numbers, but also
more complex structures such as complete distributions.  This is the
case for instance when considering the degree distribution of a graph,
or the distribution of the eigenvalues of a specific graph matrix.
Note that the reduction of a graph to a simpler space, such as that
defined by individual numbers or distributions is inherent in the
concept of \emph{graph model}:  Any graph model which would take the
whole graph and reproduce it would not be a graph model anymore.
Thus, it is crucial to the definition of a graph model that the graph
be reduced to a simpler structure.  As such, individual graph
statistics represent the simplest possible way to model a graph.  Note
also that reducing a graph to individual numbers is hardly
restrictive:  Almost all aspects of network analysis have been
expressed as graph statistics, such as the clustering coefficient for
measuring the clustering in a graph, the degree assortativity for
measuring the assortativity, etc.  Furthermore, more complex
properties such as graph spectra can themselves be reduced to
individual graph statistics, for instance by considering individual
eigenvalues or moments of given distributions.  As an example, the
classical algebraic connectivity of graphs as defined by Fiedler
\citeyearpar{b652} equals the second smallest eigenvalue of a
graph's Laplacian matrix. 

\subsection{Random Graph Models}
Synthetic graph generation is related to the concept of random graph
models.  
In the simplest formulation, a random graph model is a probability distribution over all graphs
with a given number of nodes. 
Random graph models can be specified by giving the probability of a
graph, as is done with the Erdős--Rényi model \citeyearpar{b569} and exponential
random graph models \citep{b818}, or can be specified by a randomized
algorithm, such as the preferential attachment model of Barabási and Albert \citeyearpar{b439}.
In the latter case, a random graph model can serve 
as a synthetic graph generator.  Note that a random graph model
specified as a probability distribution can be turned into a generative algorithm by
using probabilistic algorithms such as Gibbs sampling. 
In general, the purpose of random graph models is to explain the
mechanisms underlying the structure of real-world graphs, and as such
the study of random graph models is tied to the theories (sociological
or otherwise) explaining the structure of the graph.  By contrast,
other graph generators have the goal of reproducing the structures
observed in real-world graphs, without however explaining them.  
Therefore, random graph models are usually only valid on those graphs
for which the underlying theory is correct, and therefore random graph
models are as a rule not evaluated by how many different graph datasets
they are able to explain, as it is to be expected that different
real-world networks have different underlying mechanisms of evolution. 
By contrast, the algorithms presented in this paper will be formulated
as graph generators, i.e.,
with the goal to reproduce real-world graphs from many different areas,
and therefore our approach does not explain any particular underlying graph
creation mechanism, but produces synthetic graphs with 
characteristics matching many of them. 

In addition to the number of nodes, which is usually taken as fixed,
many different graph properties exist, and thus many different graph
generation algorithms have been devised, each optimizing for one or more
specific graph properties.
The first and prototypical random graph model is the one of 
Erdős and Rényi \citeyearpar{b569}, which produces  
graphs in which the number of edges  
has a given expected value.  While the
Erdős--Rényi model was not intended to generate realistic graphs, it can
be interpreted as the first of a series of graph models in which one or
more graph properties have given input values.  The Erdős--Rényi model
is conceptually and computationally simple, but produces graphs that are
highly unrealistic -- their degree distributions are Poisson
distributions and thus have a thin tail instead of a heavy one as seen
in real-world networks.
However, the average degree and characteristic distance between nodes
they produce
are usually realistic.   
In the Erdős--Rényi model, the only structural parameter apart from the
number of nodes $n$ is the individual edge probability $p$.  The
expected number of edges is then $m = p{n \choose 2}$.  Thus, the value
$p$ can be chosen specifically to generate graphs in which $m$ has a
given expected value.  The number of edges is unique in allowing such
a derivation; even random graph models that fix \emph{only} one specific
other statistic do not usually allow such a closed-form expression. 

\subsection{Preservation of Degree Distributions}
It has been observed many times that real-world networks have
power-law-like degree distributions, much different from Erdős and
Rényi's Poisson degree distributions.  Accordingly, many graph models
attempt 
to reproduce this property.  For instance, the model of 
Barabási and Albert \citeyearpar{b439}  
has been defined to grow graphs according to the rule of
preferential attachment, i.e., edges attach with preference to nodes of
high degree.  This indeed leads to graphs with power-law degree
distributions.  An alternative random graph model that reproduces realistic
degree distributions is the configuration model, also known as the
Molloy--Reed model \citeyearpar{gp15}.  This model produces a graph with the
exact same degree distribution as the input graph, but otherwise randomly
distributed edges \citep{gp54}.  A related model is that of Chung and Lu \citeyearpar{gp31},
which generates graphs in which each node has an expected degree
matching that of the input graph. Both of these algorithms have the
property that they use the full degree distribution as their parameters,
and thus their parameter space has size $O(n)$. 
Both these models fix the expected degree distribution and thus fix the
statistics that depend fully on it:  the number of $k$-stars for $k \geq 1$, which
for $k=1$ gives the number of edges. 
A generalization of these is the \textit{dK} model by Mahadevan and colleagues \citeyearpar{gp51}, which
for $k=2$ produces graphs not 
only with a given degree distribution, but with a given joint degree
distribution, i.e., the two degrees of connected nodes have the correct
expected distribution.
This additionally fixes the assortativity coefficient (i.e., the Pearson
correlation coefficient of the degree of connected nodes). 
Another generalization of the configuration model attempts to recreate
subgraph count distributions by splitting each subgraph
into \emph{hyperstubs}, i.e., individual nodes of the subgraph with
half-edges attached.  These can then be distributed among the nodes,
analogously to the configuration model \citep{gp14,gp56,gp55}.  

\subsection{Clustering}
A property not reproduced by any of the above models is clustering,
i.e., the property of real-world graphs to contain groups of nodes
well connected between each other, but less well connected to the rest
of the graph.  Clustering can be measured by the number of triangles present in a
network, or equivalently by the clustering coefficient which equals the
number of triangles normalized by the number of incident edge pairs.  Many
methods exist to produce graphs with realistic clustering.  For
instance, the Watts--Strogatz model \citeyearpar{b228} produces graphs with realistic
clustering and diameter.  As it was not created to generate 
realistic graphs \emph{per se}, it has the property that it generates unrealistic
degree distributions, like the Erdős--Rényi model.  Other models
generalize the Molloy--Reed or Chung--Lu models to incorporate
clustering to the generated graphs,  
for instance the algorithm of \citet{gp44}, and that
of \citet{gp32}.  The BTER model of Seshadhri and colleagues \citeyearpar{gp52}
is intended to reproduce not only the overall clustering coefficient,
but the distribution of the local clustering coefficient over all nodes. 
Other degree distribution-based algorithms with a clustering component
are given by Ángeles Serrano and Boguñá \citeyearpar{gp39}. 

\subsection{Matrix-based Methods}
Another class of graph properties and related algorithms are based on
algebraic graph 
theory. 
This leads to several graph models that generate a
graph's adjacency matrix from individual building blocks.  Of note in
that category is the Kronecker model \citep{b849}, which builds up an
adjacency matrix by applying the Kronecker matrix product recursively
starting with
a small initial matrix.  This model is attractive in that it allows the
size of parameter space to be varied -- it equals the number of
independent values in the model's $k \times k$ initiator matrix. 

\subsection{Exponential Random Graph Models}
Exponential random graph models \citep[ERGMs, also called p*
  models,][]{b818} are a class of random graph  
distributions based on arbitrary numerical graph statistics. ERGMs are
motivated as the maximum-entropy graph distributions with given expected
values of individual statistics. 
In theory, they can generate graphs with any given properties.  In practice,
they need very inefficient Monte-Carlo Markov chain fitting algorithms,
such that only very small graphs can be used as input \citep{b819}; large graphs
have not been generated by them. 
What is more, they often display pathological behavior, as existing
fitting algorithms will often result in extremal values of parameters,
necessitating the use of additional rules such as 
alternating families of subgraph counts \citep{b807}. 
The Erdős--Rényi model is a special case in which the number of edges
has a given expected value, and is also the only case for
undirected graphs in which a closed-form solution to the parameter
fitting problem is known.\footnote{
  A closed-form solution to this fitting problem is at
  least as complex as determining the number of graphs with a given
  number of nodes, edges, triangles, and other subgraphs.  These types
  of enumeration problems are currently out of reach of the state of the
  art, as exemplified by the fact that even the problem of counting the
  number of triangle-free graphs of a given size is highly difficult,
  and has been achieved, as of 2017, only numerically up to $n=17$
  \citep{oeis-A006785}.  
  It is also known that higher cumulants of the distribution of subgraph
  count values in random graphs go to zero in
  the large graph limit, but this does not give accurate values for
  specific sizes \citep{gp58,gp6}. 
} 

\subsection{Other Models}
Other models exist, with more or less specific goals to emulate
particular patterns of graphs or graph growth.  The Waxman model
\citeyearpar{gp27} is one based on an underlying geometry, assigning
locations in a two-dimensional space to nodes, and connecting them with
probability a function of their distance. The model is used in the
context of Internet topologies. As we want to apply the algorithms also
to networks without an underlying geometry, we will not consider it in
this paper.  What is more, the model is not amenable to fitting the
parameters to an observed graph.

\subsection{Graph Generation Strategies}
A strategy common to many graph generation algorithms consists in
creating a graph with one specific property, and then modifying the graph only in
ways that preserve this property, in order to optimize another property.
As long as such moves are possible, any graph property can be in principle recreated.
The algorithm of \citet{gp44} is a typical example: it starts with a
random network that has the correct degree distribution, and proceeds to
make switches that preserve the degree distribution but change the
number of triangles.  Thus, it is able to produce a graph with the
correct number of edges, $k$-stars and triangles.  The distribution of
certain graph statistics in such models has also been studied by Ying
and Wu \citeyearpar{gp57}.  However, other statistics such as the number of
squares will not be realistic with them.  In order to take the
number of squares into account, it would be necessary to find a series
of switches that preserve both the degree distribution and the number of
triangles -- a task that becomes intractable with an increasing number
of statistics considered.  As we will show, the proposed Guided Graph
Generator algorithm allows instead changes to any statistic at each
step, as long as the overall error (measured using all required
statistics) decreases.

Certain graph generation algorithms require as a first step to generate
a graph whose numerical properties have values in certain given ranges;
an example being the method of \citet{gp11}.  This first step is usually
non-trivial, and in fact the proposed algorithm complements these
algorithms in that they can be used as that first step.  A related but
distinct task is that of enumerating all graphs with a given exact
property \citep{gp46}; this problem only applies to very small graph
sizes and is not considered by our method.  A  
previous comparison of graph generators with respect to such strategies
is given by \citet{gp49}.

\subsection{Multi-Algorithm Methods}
Another type of graph generation model combines multiple graph
generation algorithms by choosing the appropriate one based on the requested
properties themselves.  For instance, such a method would use a
Monte-Carlo graph generator based on exponential random graphs when the
generated graphs must have a high clustering coefficient, and fall back to
a preferential attachment model when the clustering coefficient is to be
low.  An example of this type of algorithm is the GMSCN method by
\citet{gp50}. 
These types of algorithms are orthogonal to individual algorithms as evaluated
in this paper, since both can be combined.  The method for choosing an
algorithm itself must then be trained with these algorithms, leading to a
machine learning problem that necessitates a large number of input
graphs with differing characteristics; this is not necessary however for
the individual methods used in this paper. 
The approach taken by these methods contrasts with the approach taken by
our method, which is able to generate synthetic graphs whose numerical
properties cover a large range of possible values. 

\subsection{Parameter Space}
Graph models can additionally be classified by the size of their
parameter space.  Assuming that all models generate graphs with a
fixed number $n$ of nodes, graph models then differ in the
number of parameters they take:
\begin{itemize*}
\item Models such as the Erdős--Rényi model and exponential random graph
  models take a small, constant number of parameters that can be
  interpreted as graph statistics.  As an example, the Erd\H{o}s--R\'{e}nyi model
  can be understood to be parametrized by the number of nodes and edges.  The Kronecker
  model, too, takes a small constant number of parameters, i.e.\ the
  components of the initial matrix.  All these models have an
  $O(1)$-dimensional parameter space.  
\item Models that reproduce the degree distribution have the degree
  distribution as a parameter, and thus their parameter space has
  dimension $O(n)$.  This is also true for BTER, in which the clustering
  coefficient distribution serves as the parameter. 
\item Other models use a $O(n^2)$-dimensional parameter space, such as
  the method of \citet{gp43}, which starts with
  the actual 
  input graph, and modifies it iteratively.  Another class of algorithm
  with $O(n^2)$-dimensional parameter space is given by algorithms that
  use the full eigenvalue decomposition of characteristic graph matrices
  such as the Laplacian matrix \citep{gp48,gp47}. 
  These models are rare and do
  not strictly fulfill the purpose of a graph generator, since for
  instance the input graph may not be completely anonymized with them.
  Also, such algorithms have the property that they could reproduce the
  original graph's properties faithfully by not performing any changes
  at all.  Thus, the goal of these algorithms is to simultaneously
  anonymize graphs and retain their characteristics.  Due to their use of
  $O(n^2)$ parameters, their purpose is distinct from that of the
  algorithms evaluated in this paper, whose challenge lies in recreating
  the original graph's properties with only $O(1)$
  parameters. 
\end{itemize*}
The algorithm proposed in this paper has a parameter space consisting of
seven real numbers (including the number of nodes), and thus belongs to
the first class of algorithms, i.e.\ those with constant parameter size. 

\subsection{Measuring the Quality of a Graph Generation Algorithm}
As used in this article, a network statistic is a numerical measure that
characterizes a network. 
A graph generation algorithm can then be evaluated by comparing the
network statistics of the graphs it generates with the requested
values. 
In principle, we may use as an error measure any distance function based
on these.  
In the evaluation of the different algorithms, 
we will use the squared differences between the produced
statistic values and the target statistic values.  In our experiments, the
choice of whether to use the absolute value or the square did not result
in significant variation of results.  In order to avoid further
parameters in our evaluation, we thus choose to weight each statistic
equally, giving a parsimonious error measure based on the squared
differences between statistic values.  Note also that a monotonous
transformation of the error function does not result in a change of our
evaluation, or, as we will see, in the output of the proposed
algorithm. 

Given an input graph $G_0$ and a generated graph $G$,
we define the relative error with respect to a network statistic $S$ as  
\begin{align}
  E^S = \frac {S(G) - S(G_0)} {S(G_0)}
  \label{eq:errorrel}
\end{align}
in which $S(G)$ denotes the statistic value $S$ of the graph $G$. 
Based on the relative error, we then define the total error of an
algorithm at generating a graph as
\begin{align}
  E = \left[ \frac 1 {|\mathcal S|} \sum_{S \in \mathcal S} 
    (E^S)^2 \right]^{1/2},
  \label{eq:erroroverall}
\end{align}
where $\mathcal S$ is the set of considered network statistics. 
The factor $1/|\mathcal S|$ ensures that $E$ is the root mean squared
relative error of all relative errors. 
In the next section, we describe the proposed Guided Graph Generator
for minimizing 
this value $E$, and then compare the previously described baseline
algorithms with the proposed one in the subsequent section. 

\section{The Guided Graph Generator}
\label{sec:algorithm}
We are now ready to describe the Guided Graph Generator, an algorithm we
propose to generate graphs with precise values of given network
statistics. 
The algorithm is iterative; it starts with a
network that has the requested number of nodes $n$, and continues to
modify the network step by step to bring its graph statistics nearer to those
of the input graph.  
The algorithm is parametrized by a set of graph statistics $\mathcal
S$ which must be chosen when the algorithm is run -- we first describe
the algorithm in terms of that choice, and then discuss the choice in
the next section.  Note that $\mathcal S$ represents the set of
statistics, rather than statistic values; it merely encodes which
graph statistics have been chosen.  
As opposed to related iterative graph generators such as that of
\citet{gp44}, the proposed algorithm does not preserve the
properties of the graph at each step.  
Instead, we allow changes in any
property as long as the value of another property is made nearer to that
of the input graph.  

The input to the algorithm is a graph $G_0$ whose properties are
to be replicated, as well as set of
graph statistics $\mathcal S$.  As noted earlier, $\mathcal S$
contains only information about the choice of which graph statistics
are used, rather than individual values -- those can be calculated
from the given $G_0$.
The proposed algorithm works by taking as a starting point an Erdős--Rényi graph
with the correct number of nodes and edges, and then modifying the graph
iteratively until the resulting graph is as close as possible to the
target.  
To measure how close the generated graph is to the target graph, we
use the error measure $E$ as defined in
Equation~(\ref{eq:erroroverall}) in the previous section.
At each step of the iteration, we need to consider a certain number of
possible changes in the graph, and choose the change which leads to the
lowest error measure $E$.  In order to compute the changes in the statistics
efficiently, individual changes that are considered should be small,
such that the changes in the statistic values can be easily
computed. The smallest change that we can make in a graph (without
changing the node set) is to add or remove an edge.  In fact, the change
in subgraph count statistics for the addition and removal of an edge can
be expressed in terms of the immediate properties of the two involved
nodes.  Furthermore, in order to allow the algorithm to be optimized,
the computation of the changes in statistics over all changes considered
in one step should make use of common subexpressions whenever possible.
Thus, we consider  
at each step the addition and removal of edges connected to one single
node.  As we show in the next section, this leads to efficient
expressions for the change in subgraph count statistics, which leads
to efficient expressions for computing $E$ at each step. 
We note that the described algorithm, while it performs the changes
with best reduction in error at each step, is not a pure greedy algorithm, as
it does not terminate once the error cannot be further
reduced.  
The general form of the proposed Guided Graph Generator 
algorithm applying to any set of network statistics is given
in Algorithm~\ref{alg:gg}.
\begin{algorithm}[t]
  \begin{algorithmic}
    \REQUIRE a graph $G_0=(V,E_0)$ 
    \REQUIRE a set of statistics $\mathcal S$
    \REQUIRE a convergence parameter $\epsilon > 0$
    \ENSURE a graph $G=(V,E)$
    \\~\\
    \STATE $G = \textrm{ErdősRényi}(|V|, |E_0| / {|V| \choose 2})$
    \vspace{0.2cm}
    \FORALL {$S \in \mathcal S$}
    \STATE $x^S = S(G_0)$
    \STATE $y^S = S(G)$
    \ENDFOR
    \vspace{0.2cm}
    \REPEAT 
    \STATE Choose a node $u \in V$ at random
    \FORALL {$S \in \mathcal S$}
    \FORALL {$w \in V \setminus \{u\}$}
    \STATE {$\Delta^S_w = S(G \pm \{u,w\}) - S(G)$}
    \ENDFOR
    \ENDFOR
    \STATE $v = \mathrm{argmin}_{w\in V \setminus \{u\}} \sum_{S \in \mathcal S} ((y^S + \Delta^S_w - x^S) / x^S)^2$
    \STATE{$G = G \pm \{u,v\}$}
    \FORALL {$S \in \mathcal S$}
    \STATE {$y^S = y^S + \Delta^S_v$}
    \ENDFOR
    \STATE $E' = \sum_{S \in \mathcal S} ((y^S - x^S) / x^S)^2$
    \UNTIL {$E'$ has not attained a new minimum value in the last
      $(-|V| \log \epsilon)$ iterations}
  \end{algorithmic}
  \caption{
    \label{alg:gg}
    The proposed Guided Graph Generator algorithm in non-vectorized form,
    given for an arbitrary set of of graph statistics $\mathcal S$. 
  }
\end{algorithm}
We use the following notation: the function $\textrm{ErdősRényi}(n, p)$
generates an Erdős--Rényi graph with $n$ vertices and edge probability
$p$.  $G\pm \{u,v\}$ denotes the graph $G$ in which the state of the
edge $\{u,v\}$ has been switched, i.e., removed or added depending on
whether $\{u,v\}$ is present or not.  The convergence parameter
$\epsilon$ ensures that the expected ratio of nodes $u$ that were not
visited since the last new minimum value $E'$ was found equals
$\epsilon$.  In all experiments, we use a value of $\epsilon = 0.01$.

\subsection{Choice of Graph Statistics}
The Guided Graph Generator algorithm can in principle be applied to any
numerical graph statistic such as the number of triangles, the graph
diameter, or the degree assortativity.  In practice, the choice of used graph
statistics must be made such that they lead to efficient update
algorithms, and are representative of important graph characteristics.

\paragraph{Updatability.}  
To result in an efficient update algorithm, we note that properties such
as for example the graph diameter do not allow simple update expressions when
the graph is modified.  When an edge is added to a graph, we know that
the diameter cannot increase, but to compute by how much it decreases
(if at all), requires a computation almost as complex as the computation
of the diameter in the first place.  Therefore, global statistics
such as the diameter are not suited to be used in the proposed
algorithm.  
The same is true for graph statistics based on eigenvalues of
characteristic graph matrices, such as the algebraic connectivity, or
the spectral norm.
Instead, we use statistics whose change depends linearly on
local changes in the graph.  These correspond to subgraph counts,
i.e.\ the count of various subgraphs such as triangles.  As another
example, the number of 4-cliques (i.e., complete graphs $K_4$) does not allow a
simple vectorial update expression, and is therefore not used.

\paragraph{Representativity.} At the same time, the chosen statistics
should be representative of graph characteristics that are important in
practice.  For instance, the number of triangles $t$ forms the basis for the
widely used clustering coefficient ($3t/s$); the number of edges $m$
determines the graph's density ($2m/n$); and the number of squares,
being the smallest possible even cycle when multiple edges are
excluded, determines the bipartivity of the graph \citep{b456}.  The degree
distribution, itself used as a parameter for graph models, is tightly
related to the number of $k$-stars, which are related to its moments \citep{b878}.  A
$k$-star is a pattern in which a central node is connected to $k$ other
nodes.  Thus, a 2-star is a \emph{wedge}, a 3-star is a \emph{claw} and
a 4-star is a \emph{cross}.  

\paragraph{Interdependence of graph statistics.}
Certain graph statistics are related to each other in mathematically
precise ways.  For instance, the clustering coefficient $c$, defined
as the probability that two incident edges are completed by a third
edge to form a triangle, can be expressed as $c=3t/s$, where $t$ is
the number of triangles in the graph, and $s$ the number of wedges.
Thus, the clustering coefficient, while not being a subgraph count
statistic, can be recovered in graph models that optimize $t$ and
$s$.  Thus, while the algorithm presented in this paper does not
explicitly optimize for $c$, it does so implicitly because it
optimizes $t$ and $s$. 

~\\ In total, we consider six
simple possible graphs that are connected:  the edge, the wedge, the
triangle, the square, the claw and the cross.  
The complete list is given in Table~\ref{tab:patterns}.  
Note that the number of nodes
$n$ is a constant in the algorithm, and is not modified. 
These six subgraphs cover all possible
connected subgraphs up to three nodes, and those with more than three
nodes for which an update is easily expressible.  
Other graph characteristics not covered by
them such as the diameter or average path length will be subject to
experiments in Section~\ref{sec:experiments}.  

\begin{table}
  \caption{
    \label{tab:patterns}
    The subgraph patterns optimized by the Guided Graph Generator
      algorithm presented in this paper, along with a
    selection of graph properties that are represented by each subgraph
    count.  
  }
  \centering
      { \renewcommand{\arraystretch}{1.5}
        \makebox[\textwidth]{
          \scalebox{0.85}{
        \begin{tabular}{ p{1.4cm} p{6.5cm} p{5.0cm} }
          \toprule
          \textbf{Pattern} & \textbf{Statistic} & \textbf{Graph properties covered} \\
          \midrule

          \raisebox{-0.1cm}{
            \scalebox{0.5}{
              \begin{tikzpicture}
                [scale=.6,every node/.style={circle,fill=blue!80}]
                \node (n1) at (1,1) {};
                \node (n2) at (3,1) {};
                \draw (n1)--(n2);
              \end{tikzpicture}      
            }
          }
          & $m=$ volume, number of edges 
          & Density, community size \\

          \raisebox{-0.1cm}{
            \scalebox{0.5}{
              \begin{tikzpicture}
                [scale=.6,every node/.style={circle,fill=blue!80}]
                \node (n1) at (2,1.5) {};
                \node (n2) at (1,1) {};
                \node (n3) at (3,1) {};
                \draw (n1)--(n2);
                \draw (n1)--(n3);
              \end{tikzpicture}
            }
          }
          & $s=$ number of wedges/2\=/stars/2-paths
          & Degree distribution, preferential attachment \\

          \raisebox{-0.1cm}{
            \scalebox{0.5}{
              \begin{tikzpicture}
                [scale=.6,every node/.style={circle,fill=blue!80}]
                \node (n1) at (2,1) {};
                \node (n2) at (3,1) {};
                \node (n3) at (4,1) {};
                \node (n4) at (3,2) {};
                \draw (n1)--(n2);
                \draw (n2)--(n3);
                \draw (n2)--(n4);
              \end{tikzpicture}
            }
          }
          & $z=$ number of claws/3-stars 
          & Degree distribution, preferential attachment \\

          \raisebox{-0.11cm}{
            \scalebox{0.5}{
              \begin{tikzpicture}
                [scale=.6,every node/.style={circle,fill=blue!80}]
                \node (n1) at (2,2) {};
                \node (n2) at (1,1.5) {};
                \node (n3) at (1,2.5) {};
                \node (n4) at (3,2.5) {};
                \node (n5) at (3,1.5) {};
                \draw (n1)--(n2);
                \draw (n1)--(n3);
                \draw (n1)--(n4);
                \draw (n1)--(n5);
              \end{tikzpicture}
            }
          }
          & $x=$ number of crosses/4\=/stars 
          & Degree distribution, preferential attachment \\

          \raisebox{-0.1cm}{
            \scalebox{0.5}{
              \begin{tikzpicture}
                [scale=.6,every node/.style={circle,fill=blue!80}]
                \node (n1) at (2,2) {};
                \node (n2) at (1,1) {};
                \node (n3) at (3,1) {};
                \draw (n1)--(n2);
                \draw (n1)--(n3);
                \draw (n2)--(n3); 
              \end{tikzpicture}
            }
          }
          & $t=$ number of triangles/3\=/cycles/3-cliques
          & Clustering, triangle closing, analysis of triads, small-world property \\

          \raisebox{-0.1cm}{
            \scalebox{0.5}{
              \begin{tikzpicture}
                [scale=.6,every node/.style={circle,fill=blue!80}]
                \node (n1) at (1,1) {};
                \node (n2) at (1,2.0) {};
                \node (n3) at (3,2.0) {};
                \node (n4) at (3,1) {};
                \draw (n1)--(n2);
                \draw (n2)--(n3);
                \draw (n3)--(n4);
                \draw (n4)--(n1); 
              \end{tikzpicture}
            }
          }
          & $q=$ number of squares/4\=/cycles
          & Bipartivity, clustering in bipartite and nearly-bipartite graphs \\
    
          \bottomrule
        \end{tabular}
        } } }
\end{table}

\subsection{Fast Computation of $\Delta^S_w$}
\label{sec:fast-computation}
Algorithm~\ref{alg:gg} requires us to compute the difference in statistics
between the current graph $G$ and the graph $G$ with one edge 
added or removed:
\begin{align*}
  \Delta^S_w &= S(G \pm \{u,w\}) - S(G)
\end{align*}
In order for the the Guided Graph Generator to be fast, this calculation must be
performed in a vectorized way.  The existence of closed-form expressions
for $\Delta^S_w$ decides whether a particular statistic can be used in
the proposed
algorithm. 
Since $\Delta^S_w$ must be computed at each step for all nodes $w\in V$
(except for $w=v$), we derive vectorized expressions that
give a vector $\Delta^S$ containing the value $\Delta^S_w$ for all $w\in
V$.  The individual value computed for $w=v$ is then simply ignored.  

During the run of the algorithm, the graph $G$ is always represented by
its symmetric adjacency matrix $\mathbf A \in \{0,1\}^{n \times n}$.
We now give expressions for the vectors $\Delta^S$ measuring the
change in statistic $S$ expressed as functions of $\mathbf A$, for each
statistic $S \in \{m, s, z, x, t, q\}$. 
These expressions make use of the degree vector $\mathbf d$, which must be
updated along with the matrix $\mathbf A$. 
$\mathbf u \circ \mathbf v$ will denote the entrywise product between
two vectors $\mathbf u$ and $\mathbf v$, and $\mathbf A_{:u}$ the
$u$-th column of $\mathbf A$. 
In the following, we note which operations make use of a matrix-vector
product of size $n$, as these are the most expensive operations. 

\vspace{0.2cm} \noindent \textbf{Number of edges $m$.}
The number of edges will always increase or decrease by one, depending
on the previous state of the edge $\{u,w\}$.
\begin{align*}
  \Delta^m &= - 2 \mathbf A_{:u} + 1
\end{align*}

\noindent \textbf{Number of wedges $s$.}
When adding an edge, the number of wedges increases by the sum of the
degrees of the two connected nodes.  When removing an edge, the number
of wedges decreases by the sum of the degrees of the two nodes, minus
two.  
\begin{align*}
  \Delta^s &= \Delta^m \circ (\mathbf d + \mathbf d_u) + 2
  \mathbf A_{:u}
\end{align*}

\noindent \textbf{Number of claws $z$.}
When the two nodes $u$ and $w$ are not connected, the number of added
claws equals $\mathbf d_u (\mathbf d_u-1) + \mathbf d_w (\mathbf
d_w-1)$. When the two nodes are connected, the number of removed wedges
can be computed in the same way, but based on the degrees after the
removal.  
\begin{align*}
  \Delta^z &= \frac 12 \Delta^m \circ [(\mathbf d - \mathbf A_{:u})
    \circ (\mathbf d - 1 - \mathbf A_{:u}) \\
    & \qquad \qquad \,\,\,\,\, + (-\mathbf A_{:u} + \mathbf
    d_u) \circ (-\mathbf A_{:u} -1 + \mathbf d_u)]
\end{align*}

\noindent \textbf{Number of crosses $x$.}
The expression for the number of $k$-stars for higher $k$ follows the
same pattern as for $k=2$ and $k=3$.  Due to the asymmetry between the
addition and the removal of edges, the resulting expressions get
increasingly complex. 
\begin{align*}
  \Delta^x &=
  \frac 16 [
    (\mathbf d-1) \circ (\mathbf d-2) \circ (\mathbf A_{:u} \circ (-2
    \mathbf d + 3) + \mathbf d) \\
    & \qquad \;+ (\mathbf d_u-1)(\mathbf d_u-2)((3 - 2 \mathbf d_u) \mathbf A_{:u} + \mathbf d_u)
  ]
\end{align*}

\noindent \textbf{Number of triangles $t$.}
When adding an edge between two nodes, the number of added triangles
equals the number of common neighbors between the two nodes.  Likewise
when removing an edge, the number of removed triangles equals the
number of common neighbors of the two nodes.  We thus get the following
expression for the change in the number of triangles.
\begin{align*}
  \Delta^t &= (\mathbf A \mathbf A_{:u}) \circ \Delta^m
\end{align*}
We thus need to perform one sparse matrix-vector multiplication for each
iteration step. 

\vspace{0.2cm} \noindent \textbf{Number of squares $q$.}
To compute the number of squares added or removed, we count the number
of paths of length three added or removed between $u$ and $w$.  In
principle, this can be achieved by using the expression for $\Delta^t$
and multiplying $\mathbf A_{:u}$ once more by $\mathbf A$.  However,
this will also include the number of paths of length three that include
an edge $\{u,x\}$ or $\{w,x\}$ multiple times, or that include the edge
$\{u,w\}$ if it is present.  Thus, these cases must be subtracted to get
the correct number of squares added or removed.
\begin{align*}
  \Delta^q &= (\mathbf A^2 \mathbf A_{:u}) \circ \Delta^m +
  \mathbf A_{:u} \circ (\mathbf d + \mathbf d_u -1 ) 
\end{align*}
We thus need to perform two matrix-vector multiplications in this step. 

\subsection{Theoretical Runtime of the Guided Graph Generator}
\label{sec:runtime}
By computing all expressions $\Delta^S$, we 
need to perform three matrix-vector multiplications at each step.  Thus,
the runtime of each iteration step is linear in the number of current
edges $m$, and thus has runtime $O(m)$.  The number of iterations needed
for the complete algorithm cannot be derived from the algorithm itself
but must be at least $n\log \epsilon^{-1}$, and thus the total runtime of
the algorithm must be at least $O(mn \log \epsilon^{-1})$.  Under the
assumption that for large graphs, the number of edges $m$ is
proportional to $dn$, where $d$ is the size-independent average degree
of the graph, the algorithm thus has runtime quadratic in the number of
nodes $n$.  The parameter $\epsilon$ can be chosen independently of the
number of nodes, as done in the experiments, and thus does not
contribute to a further dependence of the runtime on the size $n$. 

\section{Experiments}
\label{sec:experiments}
In order to compare the common graph generation algorithms and the
proposed Guided Graph Generator, we perform three series of experiments, 
investigating their precision, scalability and
quality, as well as the empirical convergence behavior of
the proposed algorithm. 
All experiments are performed on a single machine with 72~GiB of
memory and 16 Intel Xeon X5550 processors. For the validity of the comparison,
each algorithm is run on a single core.  
We run our experiments on a set of \experimentNetworkCount network datasets,
corresponding to the \experimentNetworkCount unipartite networks with
smallest number of 
nodes available in the KONECT project\footnote{\texttt{konect.cc}}
\citep{konect}. 
The corresponding dataset names as used in the KONECT project's website
are shown in Table~\ref{tab:datasets}.  
As a running example, we also show results for a single network
specifically, the \experimentNetwork
network\footnote{\texttt{konect.cc/networks/\experimentNetworkInternal}} by
\experimentNetworkCite.  

\begin{table}
  \caption{
    \label{tab:algorithms}
    The graph generation algorithms used in our experiments, with
    their language(s) of implementation. 
  }
  \centering
  \begin{tabular*}{11cm}{ c @{\,\,\,\,} p{0.5cm} p{6.0cm} p{5.2cm} }
    \toprule
    & & \textbf{Algorithm} & \textbf{Implementation} \\
    \midrule
    \includegraphics{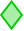} & \textsf{ER} & Erdős--Rényi \citeyearpar{b569}     & Matlab\textsuperscript{1} \\
    \includegraphics{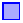} & \textsf{MR} & Molloy--Reed \citeyearpar{gp15}     & Matlab\textsuperscript{1} \\
    \includegraphics{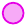} & \textsf{CL} & Chung--Lu \citeyearpar{gp31}        & Matlab\textsuperscript{1,3} \\
    \includegraphics{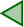} & \textsf{BB} & \citet{gp44}                        & Matlab\textsuperscript{1} \\
    \includegraphics{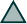} & \textsf{PP} & \citet{gp32}                        & Matlab\textsuperscript{1} \\
    \includegraphics{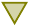} & \textsf{BT} & BTER \citep{gp52}                   & C and Matlab\textsuperscript{2} \\
    \includegraphics{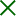} & \textsf{WS} & Watts--Strogatz \citeyearpar{b228}  & Matlab\textsuperscript{1} \\
    \includegraphics{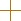} & \textsf{BA} & Barabási--Albert \citeyearpar{b439} & Matlab\textsuperscript{1} \\
    \includegraphics{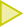} & \textsf{KR} & Kronecker \citep[$k=3$,][]{b849}    & C++\textsuperscript{2} \\
    \includegraphics{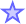} & \textsf{DK} & \textit{dK} \citep[$k=2$,][]{gp51}  & C++\textsuperscript{2} \\
    \includegraphics{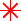} & \textsf{GG} & Guided Graph Generator              & Matlab\textsuperscript{1} \\
    \midrule
    \multicolumn{4}{l}{
      \textsuperscript{1} Custom implementation
    } \\
    \multicolumn{4}{l}{
      \textsuperscript{2} Original authors' implementation
    } \\
    \multicolumn{4}{l}{
      \textsuperscript{3} Based on implementation by \citet{gp32}
    }
  \end{tabular*}
\end{table}

In our experiments, we compare both the proposed Guided Graph Generator and
the common graph generation methods described in Section~\ref{sec:background}, as
summarized in Table~\ref{tab:algorithms}.  The methods of Erdős--Rényi,
Molloy--Reed, Chung--Lu, Barabási--Albert, and BTER are parameter-free
(for given input statistics) and are used as-is.  For the algorithm of
Bansal et al., we used an infinite number of iterations, i.e., the
algorithm always terminated properly.  For the
algorithm of Pfeiffer et al., we based the parameters on those given in
\citep{gp32}: 10,000 samples of maximal size 10,000, and
$\epsilon = 10^{-7}$.  For the Watts--Strogatz model, the parameters
were derived using the expressions given by Albert and Barabási \citeyearpar{b59}.  For the Kronecker
model, we use the implementation \emph{KronFit} \citeyearpar{kronfit} of Leskovec et al.\ with
a value of $k = 3$, i.e., a $3 \times 3$ symmetric initiator matrix with
six independent values.  We specifically do not use the newer fitting
method of Gleich and Owen \citeyearpar{b853}, as it applies only to the case $k
= 2$, which our experiments showed to produce less precise results.
For the \textit{dK} model, we used a value of $k=2$, the smallest possible.

\begin{figure}
  \centering
  \subfigure[Original network]{
    \includegraphics[width=0.40\columnwidth]{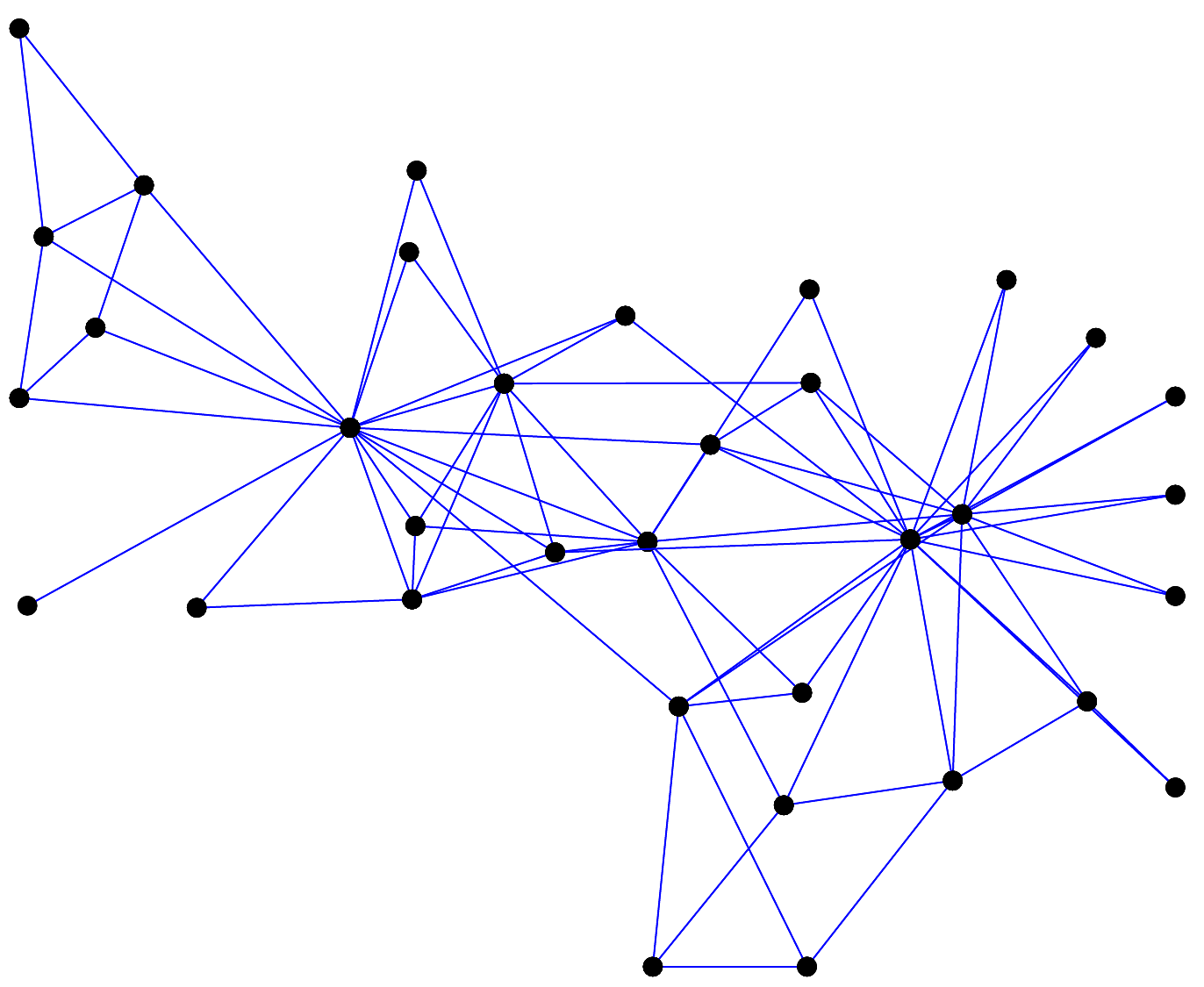}}
  \subfigure[Generated network]{
    \includegraphics[width=0.40\columnwidth]{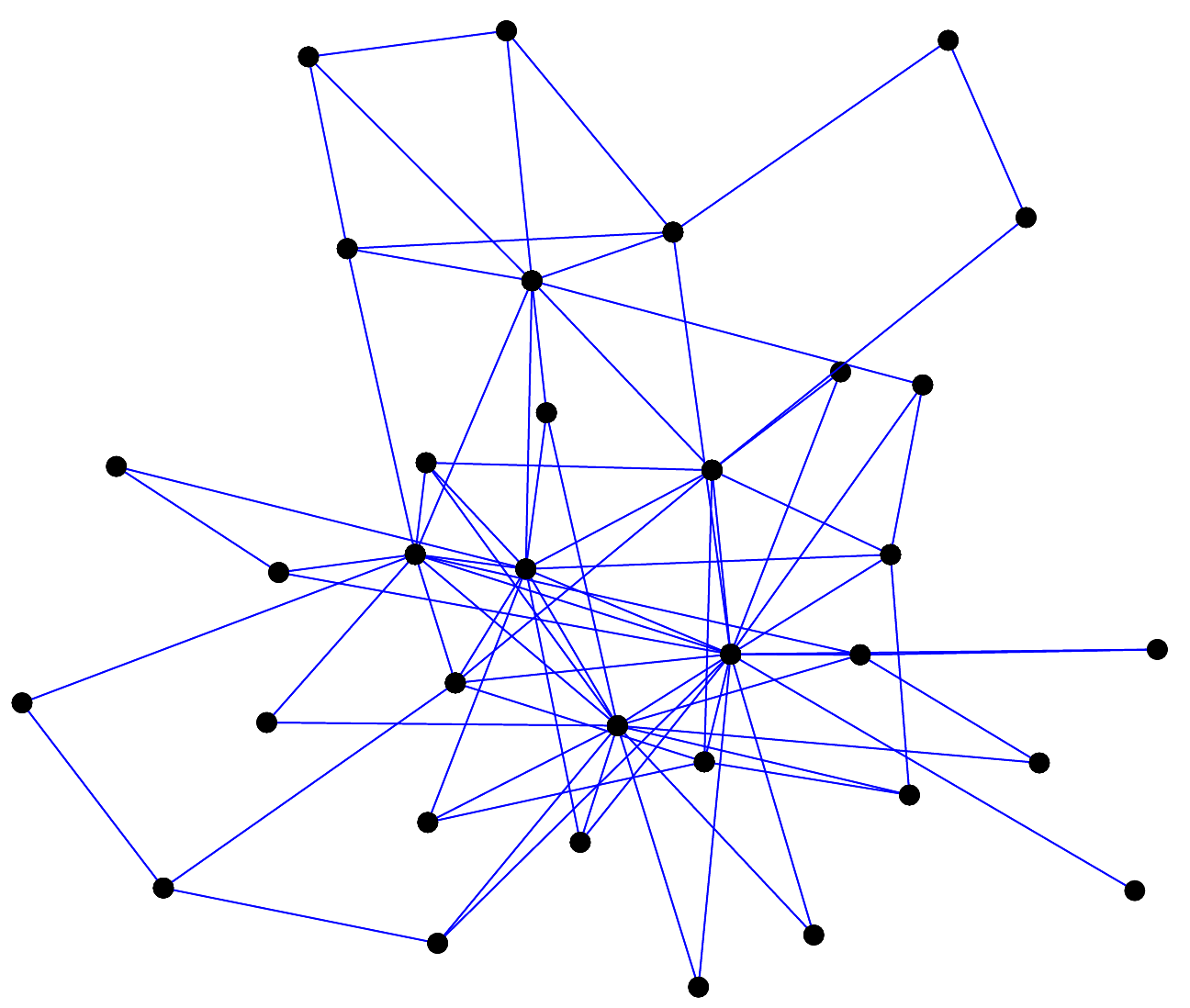}}
  \caption{
    \label{fig:experiment.precision}
    The Guided Graph Generator applied to a small network, the
    karate club network as published by \citet{konect:ucidata-zachary}:
    (a)~actual network, (b)~network generated by the proposed algorithm.
    Both networks are drawn using the algorithm of Fruchterman and
    Reingold \citeyearpar{b870}.  
  }
\end{figure}

As a small example that can be visualized, we show the graph generated
by the Guided Graph Generator algorithm, based on 
the well-known Zachary karate club
dataset\footnote{\texttt{konect.cc/networks/ucidata-zachary/}}
\citeyearpar{konect:ucidata-zachary}
in Figure~\ref{fig:experiment.precision}.  

\subsection{Precision}
\begin{figure}[t]
  \centering
  \includegraphics[width=1.0\textwidth]{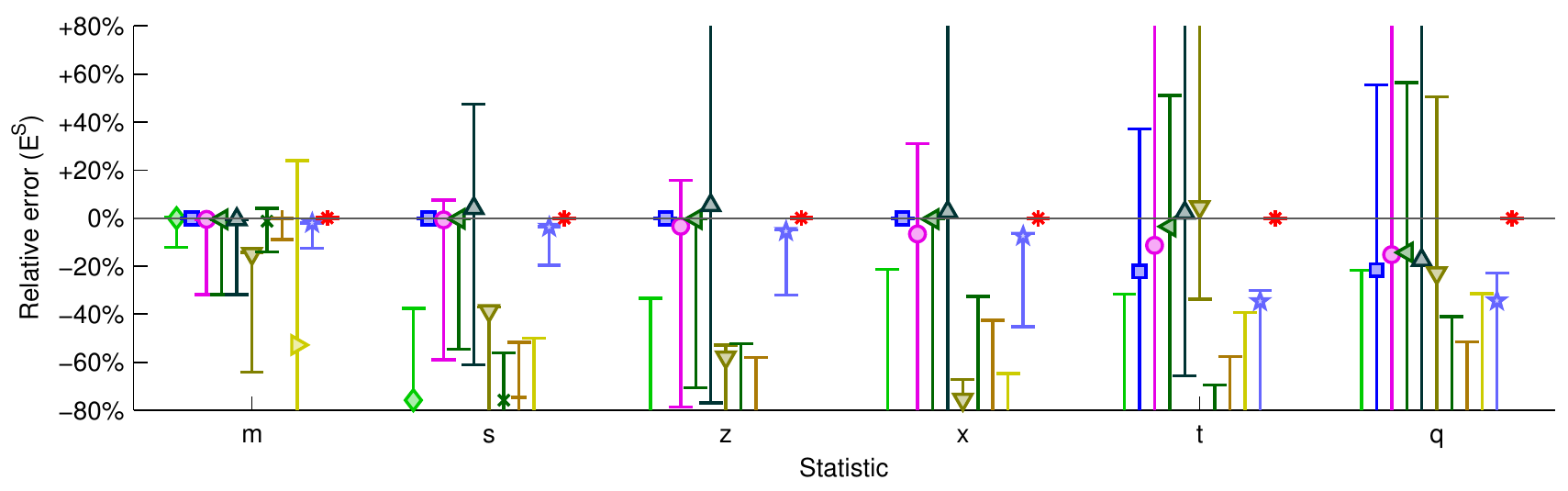}
  \includegraphics[width=1.0\textwidth]{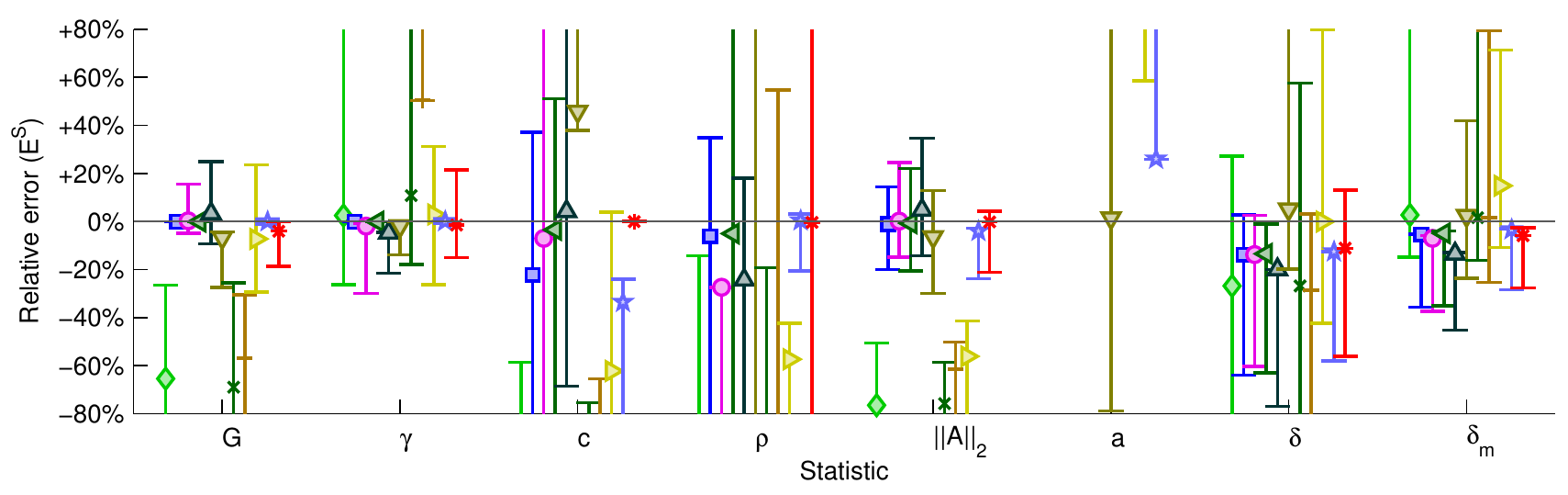}
  \scalebox{0.75}{
  \begin{tabular}{c@{\,}l @{\qquad\quad} c@{\,}l @{\qquad\quad} c@{\,}l @{\qquad\quad} c@{\,}l}
    \includegraphics{plot-svg/ER} & \textsf{Erdős--Rényi} & 
    \includegraphics{plot-svg/MR} & \textsf{Molloy--Reed} &
    \includegraphics{plot-svg/CL} & \textsf{Chung--Lu} &
    \includegraphics{plot-svg/BB} & \textsf{Bansal et al.} \\
    \includegraphics{plot-svg/PP} & \textsf{Pfeiffer et al.} &
    \includegraphics{plot-svg/BT} & \textsf{BTER} &
    \includegraphics{plot-svg/WS} & \textsf{Watts--Strogatz} &
    \includegraphics{plot-svg/BA} & \textsf{Barabási--Albert} \\
    \includegraphics{plot-svg/KR} & \textsf{Kronecker ($k=3$)} &
    \includegraphics{plot-svg/DK} & \textsf{dK ($k=2$)} &
    \includegraphics{plot-svg/GG} & \textsf{Guided Graph Generator} \\
  \end{tabular}
  }
  \caption{
    \label{fig:precision3}
    Average relative error $E^S$ of each algorithm for each statistic
    $S$.  The plots show the median relative error, as well as the
    10\textsuperscript{th} and 90\textsuperscript{th} percentiles
    computed over the \experimentNetworkCount networks in our
    experiments.  The top row shows the statistics that are optimized by
    the Guided Graph Generator; the bottom row shows other statistics.
    The legend for the individual graph statistics is given in Tables
    \ref{tab:patterns} and \ref{tab:statistics}.  
  }
\end{figure}

In the first experiment, we ask to what precision the algorithms are
able to generate graphs.  We verify both the precision in terms of the
six subgraph count statistics optimized by the proposed algorithm as shown in
Table~\ref{tab:patterns}, as well as in terms of other common graph
statistics as given in Table~\ref{tab:statistics}.  While we expect the
proposed algorithm to be precise in terms of the optimized statistics,
there is \emph{a~priori} no reason to believe it should perform well for
the other statistics.  Table~\ref{tab:precision2} shows the statistics
of the graphs generated with this and the other algorithms for the
example of the \experimentNetwork network as published by
\experimentNetworkCite.  To measure the precision of the algorithm over
all \experimentNetworkCount investigated networks, we compute the
average relative error of the proposed and other algorithms over all of
them.  Defining the average relative error of an algorithm with respect
to a given statistic $S$ as the average of the relative errors $E^S$
from Equation~(\ref{eq:errorrel}) computed over all considered networks,
Figure~\ref{fig:precision3} shows the average relative error for all
algorithms.

\begin{table}[t]
  \caption[*]{
    \label{tab:statistics}
    Additional network statistics that are not explicitly optimized by the Guided Graph
    Generator algorithm, but used in our evaluation. 
    When not noted otherwise, the
    definitions of the statistics we use are those given in the Handbook
    of the KONECT project
    \citep{konect:handbook}.  We also give network analysis methods
    in which these statistics are used.
  }
  \centering
  \scalebox{0.9}{
  \renewcommand{\arraystretch}{1.5}
  \setlength\tabcolsep{0.1cm}
  \begin{tabular}{ c p{5.9cm} @{\hspace{0.5cm}} p{4.8cm} }
    \toprule
    \textbf{Statistic} & \textbf{Definition} & \textbf{Analysis methods} \\
    \midrule
    $G$ & Gini coefficient \citep{kunegis:power-law} & Equality of the
    degree distribution \\
    $\gamma$ & Power law exponent \citep[Eq. 5]{b408} & Scale-free
    network analysis \\
    $c$ & Clustering coefficient ($=3t/s$) \citep{b736} & Small-world analysis \\
    $\rho$ & Assortativity, i.e.\ Pearson correlation coefficient
    between the degree of connected nodes \citep{b854} & Homophily analysis \\
    $\| \mathbf A \|_2$ & Spectral norm, i.e.\ largest
    absolute eigenvalue of the adjacency matrix $\mathbf A$ & Network
    growth analysis \\
    $a$ & Algebraic connectivity, i.e.\ smallest nonzero eigenvalue of
    the Laplacian matrix  \citep{b652} & Connectivity analysis \\
    $\delta$ & Graph diameter \citep{b779} & Small-world analysis, connectivity analysis \\
    $\delta_{\mathrm m}$ & Average distance between two nodes
    \citep{b779}
    & Small-world analysis, connectivity analysis \\
    \bottomrule
  \end{tabular}
  }
\end{table}

From the case of the \experimentNetwork network, and from the results
over all networks, we make the following observations. 
The Guided Graph Generator reproduces most statistics with higher precision
than the other algorithms, even statistics that are not optimized by
it. The exception is the algebraic connectivity, which is not well
reproduced by any algorithm, but still better reproduced by the \textit{dK}
model, and the diameter, which is better reproduced by the Kronecker
model.  We note that the average distance however is better reproduced
by the Guided Graph Generator.  In particular, the proposed algorithm generates graphs with
better fitting values of the number of triangles $t$ and squares $q$,
which are important for clustering and bipartivity characteristics. 
Also, the fact that the errors in the different statistics are so low
for the Guided Graph Generator is an indication that the algorithm does
not get stuck in any local optimum, i.e., it actually reaches an optimum
very near to the requested values -- the small derivation from the
requested values can then be explained by combinatorial arguments. 

While for individual statistics individual
algorithms are as good as ours, this is not true for all statistics
combined, including the non-optimized ones.  The overall performance
is better for more statistics with the proposed algorithm. 
In particular, we make the following observations:

\vspace{0.2cm} \noindent The \textbf{number of edges} $m$ is matched by
all algorithms, except the Kronecker model, which produces exact values
only in powers of the base matrix size.

\vspace{0.2cm} \noindent The \textbf{number of wedges} $s$, as an
indicator of the inequality of the degree distribution, is matched
approximately by all except the Erdős--Rényi, Watts--Strogatz and
Barabási--Albert models.  For the latter one, this is unexpected, as
that algorithm is intended to produce realistic degree distributions,
but can be explained by the lack of methods to adjust the algorithm to a
given number of wedges.
The number of claws $z$ and crosses $x$ follow similar patterns as the
number of wedges.

\vspace{0.2cm} \noindent The \textbf{number of triangles} $t$ is badly
reproduced by most algorithms.  The three classical algorithms of Erdős
and Rényi, Watts and Strogatz, and Barabási and Albert without
consideration of clustering produce graphs with orders of magnitudes too
few triangles.  The other algorithms, which do consider clustering,
produce numbers of triangles within a factor of two of the correct
value.  All algorithms expect BTER, the algorithms of Bansal et al., 
Pfeiffer et al., and the Guided Graph Generator produce graphs with too
few triangles.  The clustering coefficient $c$ shows similar behavior.

\vspace{0.2cm} \noindent The \textbf{number of squares} $q$ is matched
well only by the Guided Graph Generator.  It is thus the only graph
generator that takes into account bipartivity among those tested. 

\vspace{0.2cm} \noindent The \textbf{Gini coefficient} $G$ is matched
well by all algorithms based on the degree distribution, as expected.
The classical algorithms of Erdős--Rényi, Watts--Strogatz and
Barabási--Albert do not match it.  Other algorithms match it reasonably
well, and the Guided Graph Generator matches it very well.

\vspace{0.2cm} \noindent The \textbf{power law exponent} $\gamma$ does
not have a large range of values in the generated graphs, and thus most
algorithms match it well.  A notable exception is the Barabási--Albert
model, which produces values that are too high by up to 50\%, consistent
with the fact that the actual values of the exponent seen in real graphs
are for the most part smaller than the Barabási--Albert model's
theoretical value of three. 

\vspace{0.2cm} \noindent The \textbf{degree assortativity} $\rho$ is
only matched well by the \textit{dK} model, which includes the joint
degree distribution as a parameter, and thus fixes $\rho$.
We note that it may be possible to achieve a much more precise value of
the degree assortativity for the Guided Graph Generator if it were
possible to include paths of length three as a pattern, as the number of
these subgraphs is related to the sum of degrees of a node's neighbors.
The number of 3-paths could however not be used as it is too expensive
to keep up to date in the algorithm. 

\vspace{0.2cm} \noindent The \textbf{spectral norm} $\| \mathbf A \|_2$
is matched well by most algorithms, with the notable exception of the
Kronecker model.  This is somewhat surprising as the Kronecker model is
defined using matrix multiplication. 

\vspace{0.2cm} \noindent The \textbf{algebraic connectivity} $a$ is
badly matched by all algorithms.  As
the algebraic connectivity characterizes the graph globally, we should
expect models that generate specific structures for the graph as a
whole to match it.  This is the case for the Kronecker model, even though its
resulting algebraic connectivity does not match that of the input
graphs.

\vspace{0.2cm} \noindent The \textbf{average distance} $\delta_{\mathrm
  m}$ is matched reasonably well by all algorithms.  In particular, even
the Erdős--Rényi model matches it.  Most algorithms produce too small
diameters $\delta$ however, except for the Kronecker model.

\vspace{0.2cm} We conclude from these observations that a precise matching of a
graph's features is complex:  Even algorithms designed to reproduce a
certain feature often fall only very approximately near the correct
value.  This is due to various reasons.  For the Kronecker algorithm,
this is due to the fact that only graphs whose size is a power of the
initial matrix can be generated, and thus a downsampling step would be needed
afterwards, complicating matters.  Other methods fail because of too hard 
constraints -- the algorithm of Bansal et al.\ for instance fails to
generate graphs with the required amount of triangles, even though it is
designed 
to do so, because preserving the exact degree distribution is too
strong a constraint.  We also observe the pattern that algorithms intended
to reproduce one feature exactly often fail greatly at reproducing
other features, 
to the point where a simpler algorithm would be better. For instance, the
Watts--Strogatz model was specially constructed to produce realistic
diameters and clustering, but produces unrealistic degree
distributions. 

\begin{table}
  \caption{
    \label{tab:datasets}
    The \experimentNetworkCount datasets used in our experiments.  The
    names correspond to the names used by the KONECT project
    \citep{konect:handbook} given on \texttt{konect.cc}. 
  }
  \centering
  \begin{tabular}{p{0.95\textwidth}}
    \toprule
    {\footnotesize \texttt{ucidata-gama ucidata-zachary mit adjnoun\_adjacency sociopatterns-hypertext foodweb-baydry foodweb-baywet radoslaw\_email contact sociopatterns-infectious arenas-meta arenas-email subelj\_euroroad opsahl-usairport opsahl-ucsocial ego-facebook opsahl-openflights opsahl-powergrid subelj\_jung-j subelj\_jdk as20000102 advogato elec lasagne-frenchbook arenas-pgp dblp-cite lasagne-spanishbook cfinder\_google ca-AstroPh eat subelj\_cora ego-twitter ego-gplus as-caida20071105 hep-th-citations munmun\_digg\_reply
} } \\
    \bottomrule
  \end{tabular}
\end{table}

\begin{table}
  \caption{
    \label{tab:precision2}
    Statistics of graphs generated to be similar to the
    \experimentNetwork network by \experimentNetworkCite\ with $n = \experimentNetworkSize$.  
    $G_0$ denotes the properties of the actual graph. 
    See Tables \ref{tab:patterns}, \ref{tab:algorithms}, and \ref{tab:statistics} for
    the list of graph generators as well as network statistics. 
  }
  \centering
  \makebox[\textwidth]{
    \scalebox{0.85}{
      \setlength\tabcolsep{0.16cm}
\begin{tabular}{ l r r r r r r r r r r r r r r }
\toprule
 & $\boldsymbol{m}$  & $\boldsymbol{s}$  & $\boldsymbol{z}$  & $\boldsymbol{x}$  & $\boldsymbol{t}$  & $\boldsymbol{q}$  & $\boldsymbol{G}$  & $\boldsymbol{\gamma}$  & $\boldsymbol{c}$  & $\boldsymbol{\rho}$  & $\boldsymbol{||A||_2}$  & $\boldsymbol{a}$  & $\boldsymbol{\delta}$  & $\boldsymbol{\delta_m}$ \\
\midrule
$G_0$ & $24{,}316$  & $434{,}797$  & $7{,}501{,}208$  & $180{,}494{,}388$  & $54{,}788$  & $1{,}010{,}957$  & $59\%$  & $2.11$  & $37.8\%$  & $0.238$  & $42.4$  & $0.011$  & $24$  & $7.48$   \\
\midrule
  \textsf{ER}  & $24{,}334$  & $110{,}717$  & $167{,}083$  & $186{,}911$  & $13$  & $51$  & $25\%$  & $1.71$  & $0.0\%$  & $0.005$  & $5.77$  & $0.218$  & $12$  & $6.27$  \\
  \textsf{MR}  & $24{,}316$  & $434{,}797$  & $7{,}501{,}208$  & $180{,}494{,}388$  & $880$  & $11{,}541$  & $59\%$  & $2.11$  & $0.6\%$  & $-0.017$  & $20.2$  & $0.066$  & $12$  & $4.70$  \\
  \textsf{CL}  & $24{,}316$  & $465{,}769$  & $8{,}376{,}222$  & $212{,}662{,}941$  & $1{,}115$  & $15{,}287$  & $57\%$  & $1.88$  & $0.7\%$  & $-0.009$  & $21.5$  & $0.115$  & $12$  & $4.41$  \\
  \textsf{BB}  & $24{,}316$  & $434{,}797$  & $7{,}501{,}208$  & $180{,}494{,}388$  & $875$  & $10{,}969$  & $59\%$  & $2.11$  & $0.6\%$  & $-0.021$  & $20.1$  & $0.067$  & $13$  & $4.69$  \\
  \textsf{PP}  & $24{,}316$  & $517{,}195$  & $9{,}275{,}916$  & $232{,}859{,}495$  & $6{,}795$  & $28{,}076$  & $57\%$  & $1.75$  & $3.9\%$  & $-0.017$  & $23.3$  & $0.104$  & $11$  & $4.14$  \\
  \textsf{BT}  & $24{,}160$  & $419{,}760$  & $6{,}401{,}963$  & $131{,}265{,}941$  & $48{,}281$  & $768{,}150$  & $56\%$  & $1.89$  & $34.5\%$  & $0.316$  & $39.6$  & $0.025$  & $17$  & $5.39$  \\
  \textsf{WS}  & $21{,}360$  & $73{,}999$  & $73{,}727$  & $47{,}782$  & $4{,}234$  & $3{,}054$  & $19\%$  & $1.76$  & $17.2\%$  & $0.050$  & $4.74$  & $0.076$  & $18$  & $8.40$  \\
  \textsf{BA}  & $24{,}264$  & $113{,}844$  & $198{,}710$  & $296{,}103$  & $59$  & $173$  & $27\%$  & $2.42$  & $0.2\%$  & $0.165$  & $7.39$  & $0.525$  & $10$  & $6.25$  \\
  \textsf{KR}  & $12{,}422$  & $107{,}857$  & $895{,}235$  & $9{,}793{,}414$  & $83$  & $470$  & $49\%$  & $2.16$  & $0.2\%$  & $-0.056$  & $11.0$  & $0.067$  & $15$  & $5.49$  \\
  \textsf{DK}  & $23{,}206$  & $420{,}091$  & $7{,}214{,}212$  & $172{,}216{,}188$  & $6{,}306$  & $138{,}845$  & $60\%$  & $2.10$  & $4.5\%$  & $0.232$  & $33.8$  & $0.024$  & $19$  & $5.37$  \\
  \textsf{GG}  & $24{,}317$  & $434{,}799$  & $7{,}501{,}218$  & $180{,}494{,}690$  & $54{,}788$  & $1{,}010{,}955$  & $60\%$  & $2.06$  & $37.8\%$  & $0.032$  & $33.4$  & $0.003$  & $28$  & $8.83$  \\
\bottomrule
\end{tabular}

    }
  }
\end{table}

\subsection{Scalability of the Guided Graph Generator}
We have shown in Section~\ref{sec:runtime} that each iteration step of
the Guided Graph Generator has
a runtime of $O(m)$, where $m$ is the number of edges in the graph.  The
number of iterations needed for the proposed algorithm 
to converge cannot be deduced theoretically, but a simple heuristic
dictates that if all nodes should be visited, then the number of
iterations will also be linear in $n$, giving a total runtime of
$O(mn)$.  

\begin{figure}
  \centering
  \subfigure{
    \includegraphics[width=0.30\columnwidth]{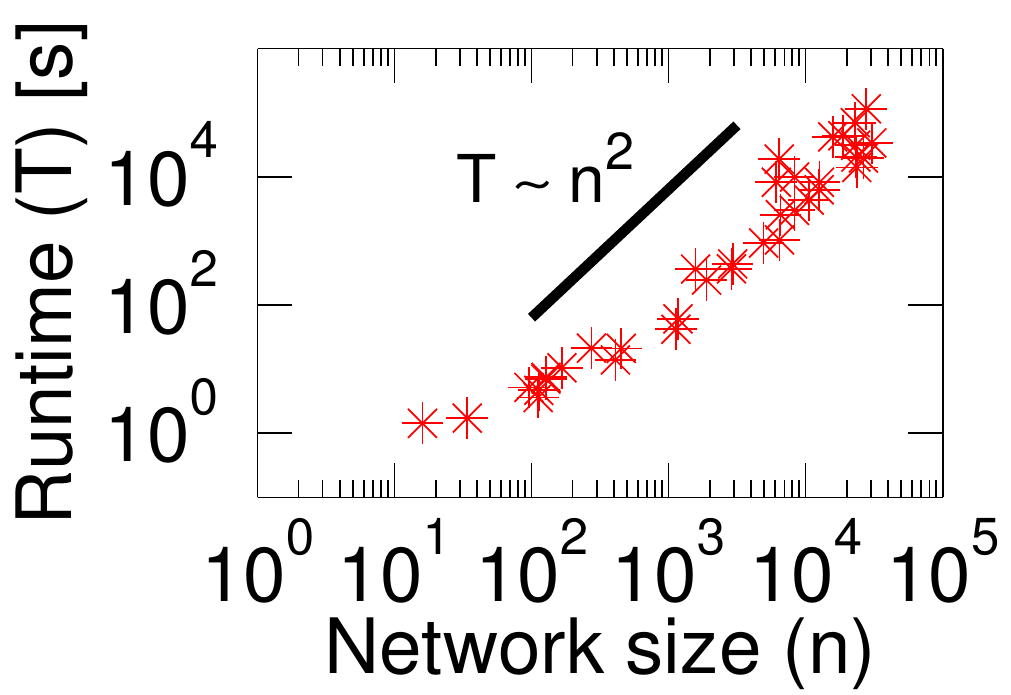}}
  \subfigure{
    \includegraphics[width=0.30\columnwidth]{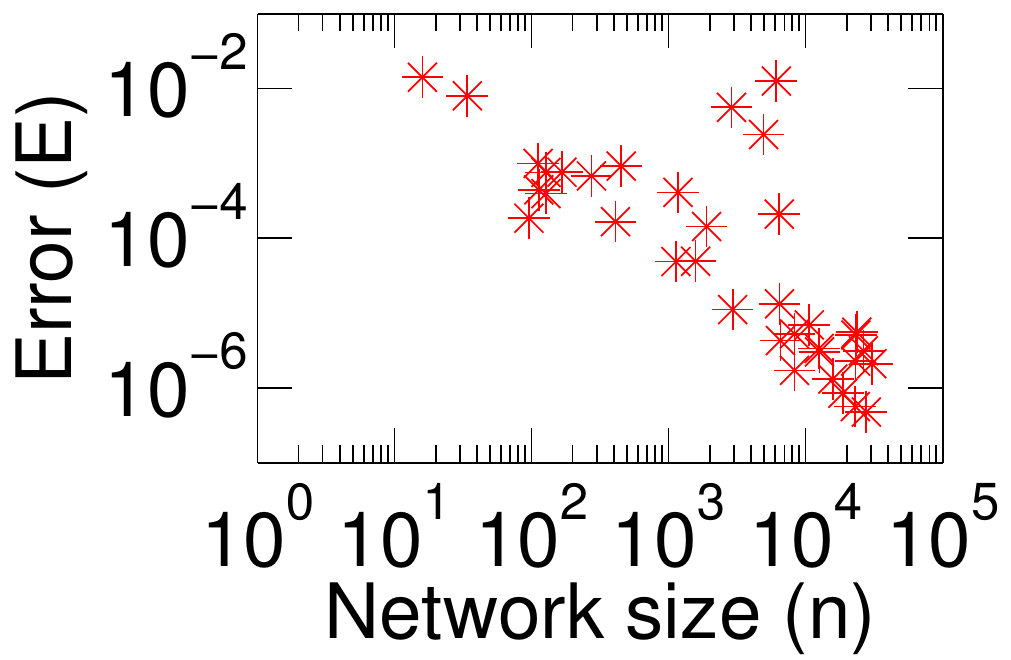}}
  \caption{
    \label{fig:experiment.scalability}
    The error and runtime of the Guided Graph Generator in function of network size,
    i.e.\ the number $n=|V|$ of nodes. 
  }
\end{figure}

For all networks that we use in this paper, the Guided Graph Generator
as implemented in the Matlab programming language took at most
28 hours to complete, and used no more than 5~GiB of memory.  For
comparison, fitting the Kronecker model is said to take from 24 to 48
hours and 32 GiB of RAM for networks with 
200,000--300,000 nodes \citep{gp49}.  In order to measure the runtime's
exponent as a function of network size, we show in
Figure~\ref{fig:experiment.scalability} the runtime and network size, on
a doubly logarithmic plot.  The results are consistent with a runtime quadratic in
network size. The same experiments with the other methods (not shown)
resulted in similarly quadratic runtimes, except for the Kronecker
model, whose fitting algorithm was slow with small networks, but not
slower for larger networks, making it impossible to make any statement
about its asymptotic runtime as a function of network size.

\subsection{Analysis of Characteristic Network Distributions}
Network statistics do not uniquely determine a graph, and we must thus
ask whether the consideration of a few network statistics is sufficient
to claim that a generated network is realistic.  In order to do this, we
consider multiple properties of networks that are not represented by
individual numbers but by a whole distribution of values:  
the degree distribution, the clustering coefficient distribution, the
distance distribution, and the spectral distribution.  
All plots shown in this section are for the \experimentNetwork network. 

\begin{figure}
  \centering
  \includegraphics[width=0.55\columnwidth]{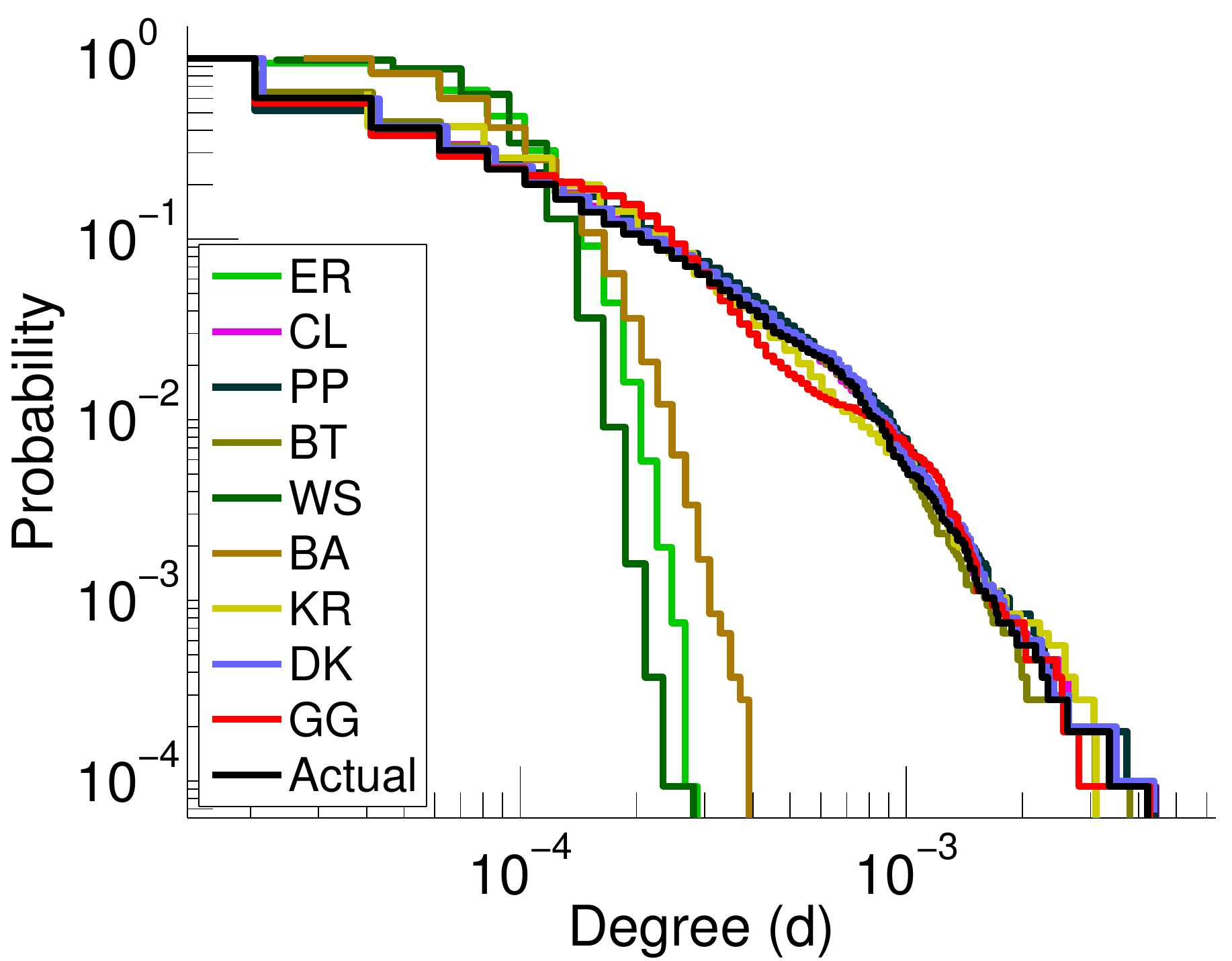}
  \caption{
    \label{fig:experiment.qualitative.degree}
    Comparison of the complementary cumulative degree distributions
    generated by the different 
    models for the \experimentNetwork network. 
  }
\end{figure}

\vspace{0.2cm} \noindent \textbf{Degree Distribution.}
The degree of a node is the number of its neighbors, and accordingly the
degree distribution of a network can be considered.  Real-world networks
have been observed, many times, to have degree distributions with
power-law tails \cite[see e.g.][]{b462}.   This is as opposed to models such as Erdős--Rényi
random graphs, which have Poisson degree distributions. 
Figure~\ref{fig:experiment.qualitative.degree} shows the degree
distributions of the graphs generated by the various methods, excluding
those methods that take the degree distribution as input
and thus generate the exact correct degree distribution. 
We observe that all methods except those of Erdős--Rényi, Watts--Strogatz and
Barabási--Albert produce well-fitting degree distributions. A further
observation can be made about the Guided Graph Generator:  Its degree distribution is
not as smooth as the original one, but deviates slightly in alternate
directions.  We explain this by the fact that the proposed algorithm takes not
the full degree distribution as input, but only the number of edges,
wedges, claws and crosses, i.e.\ the number of $k$-stars
for $k=1,2,3,4$, where we interpret an edge as a 1-star.  Since the numbers
of $k$-stars are related to the $k$-th moments of
the degree distribution\footnote{The difference is that the moments are
  defined as sum of powers of node degrees, while the number of
  $k$-stars equals the sum of falling powers of node degrees.}, the proposed
algorithm generates graphs whose degree 
distributions are correct up to these modified moments. 
We also note the incidental similarity of graphs generated by the proposed
algorithm to Kronecker graphs, as those too tend to have unbalanced degree
distribution, containing rough steps. 

\begin{figure}
  \centering
  \includegraphics[width=0.55\columnwidth]{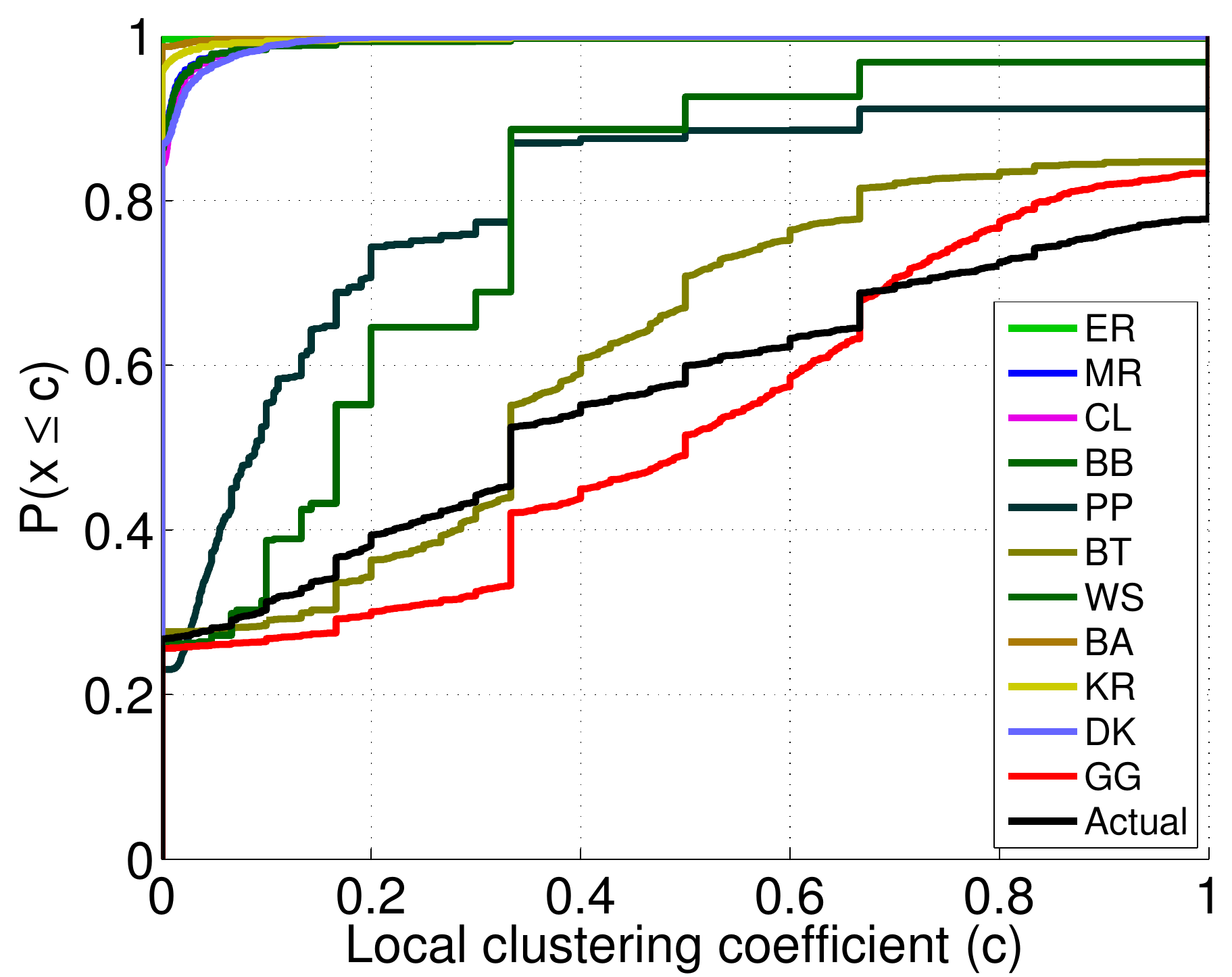}
  \caption{
    \label{fig:experiment.qualitative.cluscod}
    Comparison of the local clustering coefficient distributions generated by
    the different models for the \experimentNetwork network. 
  }
\end{figure}

\vspace{0.2cm} \noindent \textbf{Clustering Coefficient Distribution.}
The clustering coefficient $c$ as defined previously is a global
characteristic, denoting the probability that two nodes with a common
neighbor are connected.  This measure of clustering can also be applied
to individual nodes, giving the local clustering coefficient, i.e.\ the
probability that two neighbors of a given node are connected. The
distribution of the local clustering coefficient over all nodes then
gives a network's clustering coefficient distribution. 
Figure~\ref{fig:experiment.qualitative.cluscod} shows the clustering
coefficient distribution of the networks generated by each algorithm. 
We observe that the clustering coefficient distributions of the different graph
models vary wildly:  Models that do not take into account clustering
have, as expected, almost only nodes with a local clustering coefficient
near zero.  The algorithm of Bansal et al., as well as that of Pfeiffer
et al.\ produce slightly more correct clustering coefficient
distributions. 
Finally, the best clustering coefficient distribution are generated by
BTER and by the Guided Graph Generator.  For BTER, this is to be excepted, as BTER takes
the local clustering coefficient distribution as input. 

\begin{figure}
  \centering
  \includegraphics[width=0.50\columnwidth]{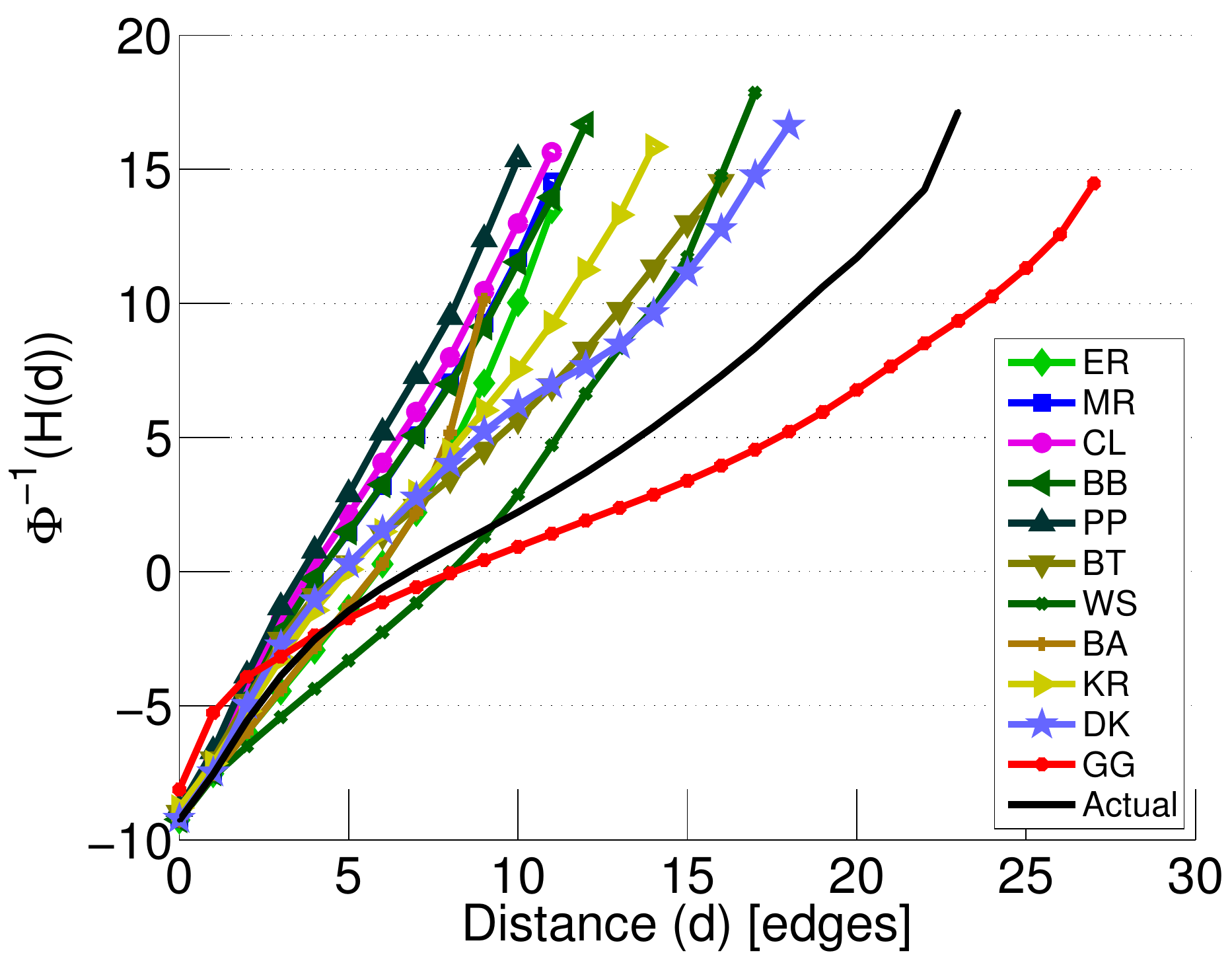}
  \caption{
    \label{fig:experiment.qualitative.hopdistr}
    Comparison of the distance distributions generated by
    the different models for the \experimentNetwork network. 
  }
\end{figure}

\vspace{0.2cm} \noindent \textbf{Distance Distribution.}
The distance between two nodes of a graph is defined as the minimum number of
edges needed to reach one node from the other.  
Distances determine the
dynamics of communication within a network, and are therefore of
importance for many types of networks. 
The distance distribution is the distribution of distance values over all node
pairs. Thus, the distance distribution extends the average path
length $\delta_{\mathrm m}$ and the maximal path length $\delta$ (the
diameter) to give information about the distribution over all path lengths. 
In order to plot the distance distribution, we use its cumulative
distribution function $H(d)$, i.e.,
the proportion of all node pairs that are reachable in at most $d$
edges.  The resulting plots in
Figure~\ref{fig:experiment.qualitative.hopdistr} show this function
$H(d)$ on an inverse logistic scale, i.e.\ we show $\Phi^{-1}(H(d)) =
\ln(H(d) / (1 - H(d)))$ on
the Y axis, where $\Phi(x) = 1/(1+e^{-x})$ is the logistic function. 
The reason for choosing a logistic scale is to give particular attention
to the tails of the distribution, as those are otherwise not captured by
the average path length. 
We choose the cumulative distribution for visualization as it is related
to the Kolmogorov--Smirnov test (and thus the Kolmogorov--Smirnov
distance) which measures the similarity of two distributions. 
As none of the tested algorithms optimizes directly for distances in the
graphs, none produces a particularly well matching distance
distribution.  The Guided Graph Generator, too, does not produce a distance
distribution that matches the given network with precision.  As noted before
though, we have identified that Kronecker graphs match original graphs
well in their diameter, i.e.\ the maximum of the distance distribution,
and the proposed algorithm matches original graphs well in the average of the
distance distribution.  It remains thus an open problem to define a
graph model that reproduces the distance distribution accurately. 

\vspace{0.2cm} \noindent \textbf{Spectral Distribution.}  
An important characterization of a graph is in terms of the spectrum of
its adjacency matrix, which captures information about the number of
cycles of different lengths it contains.  We consider the distribution of
the eigenvalues of a graph's normalized adjacency matrix $\mathbf N$.
This matrix is an $n \times n$ matrix, defined by $\mathbf N_{uv} =
1/\sqrt{\mathbf d_u \mathbf d_v}$ when $u$ and $v$ are connected, and
$\mathbf N_{uv}=0$ otherwise.  $\mathbf N$ is symmetric, and its real
eigenvalues lie in the interval $[-1, +1]$.  The set of eigenvalues of
the matrix $\mathbf N$ encodes information about the number of cycles of
length $k$ for all $k \geq 0$, in that the $k$-th
moment of the eigenvalues equals the probability of a random walk of
length $k$ to return to its starting point.  Thus, the spectral
distribution is an extension of the number of triangles $t$ and squares
$q$ to longer cycles, and comparing the spectrum of $\mathbf N$ serves
as a test of the accuracy for graph generators to generate graphs in
which the number of higher-order cycles is realistic.

\begin{figure}
  \centering
  \includegraphics[width=0.5\columnwidth]{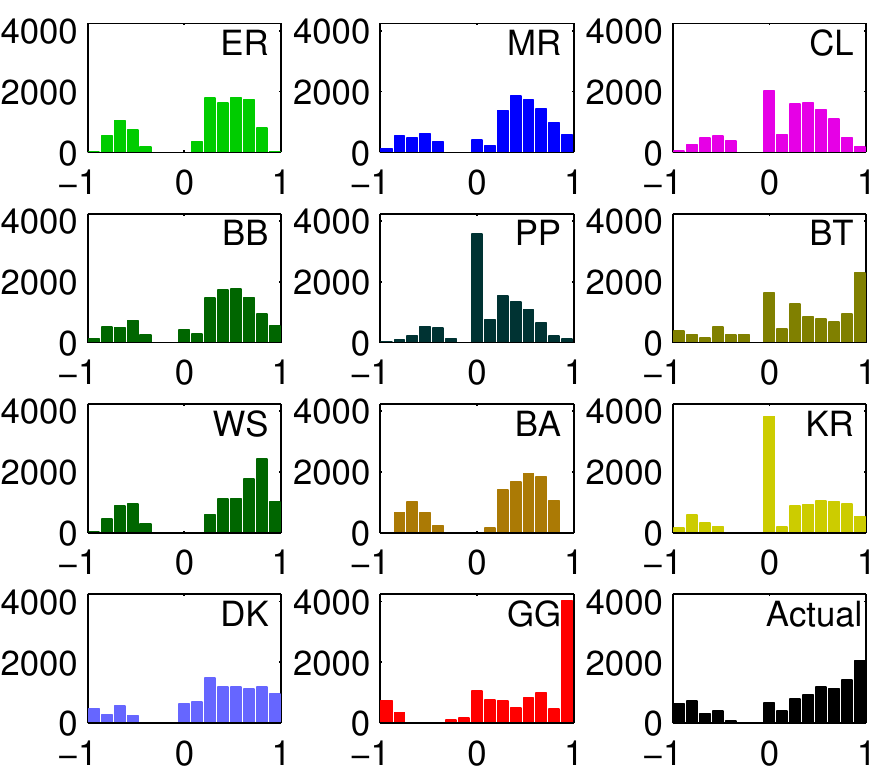}
  \caption{
    \label{fig:experiment.qualitative.distr}
    Comparison of the normalized spectral distributions generated by
    the different models for the \experimentNetwork network, based on
    the eigenvalues of the normalized adjacency matrix $\mathbf N$ of
    the networks. 
  }
\end{figure}

Figure~\ref{fig:experiment.qualitative.distr} shows the spectral
distribution of the networks generated by the different methods.  We
observe that none of the algorithms reproduces the spectral distribution
precisely.  The methods that come the closest are BTER and \textit{dK}.
A few observations can be made from the specific spectra of individual
models: The Kronecker model, as well as the model of Pfeiffer et
al.\ have a large number of near-zero eigenvalues; this indicates that
they produce graphs with many unconnected or badly-connected vertices.
The Guided Graph Generator algorithm produces a graph with many
eigenvalues equal to one, indicating that it creates too many small non-empty
clusters unconnected to the rest of the network.  Both of these are
errors in the reconstruction, as the original graph does not have these
properties.

\subsection{Convergence of the Guided Graph Generator}
The proposed Guided Graph Generator algorithm is greedy-like in that
it performs only local optimizations by adding and removing individual
edges.  Thus, there is no theoretical guarantee that the algorithm
cannot get stuck in a local optimum.  As an empirical test of the
behavior of the algorithm in real-world cases, we may inspect
Figure~\ref{fig:experiment.scalability}:  The error $E$ as a
function of the network size $n$ is decreasing, indicating that
the generated graphs are reasonably close to the requested target
values, and that the relative error is smaller for larger graphs.
Thus, while it would be conceivable that the algorithm gets stuck in a
local optimum for certain ranges of input subgraph counts, this
appears to not be the case.  We can conclude from this that the
property of a graph that makes the Guided Graph Generator get stuck is
a very rare property in real-world graphs. 

\begin{figure}
  \centering
  \includegraphics[width=0.7\columnwidth]{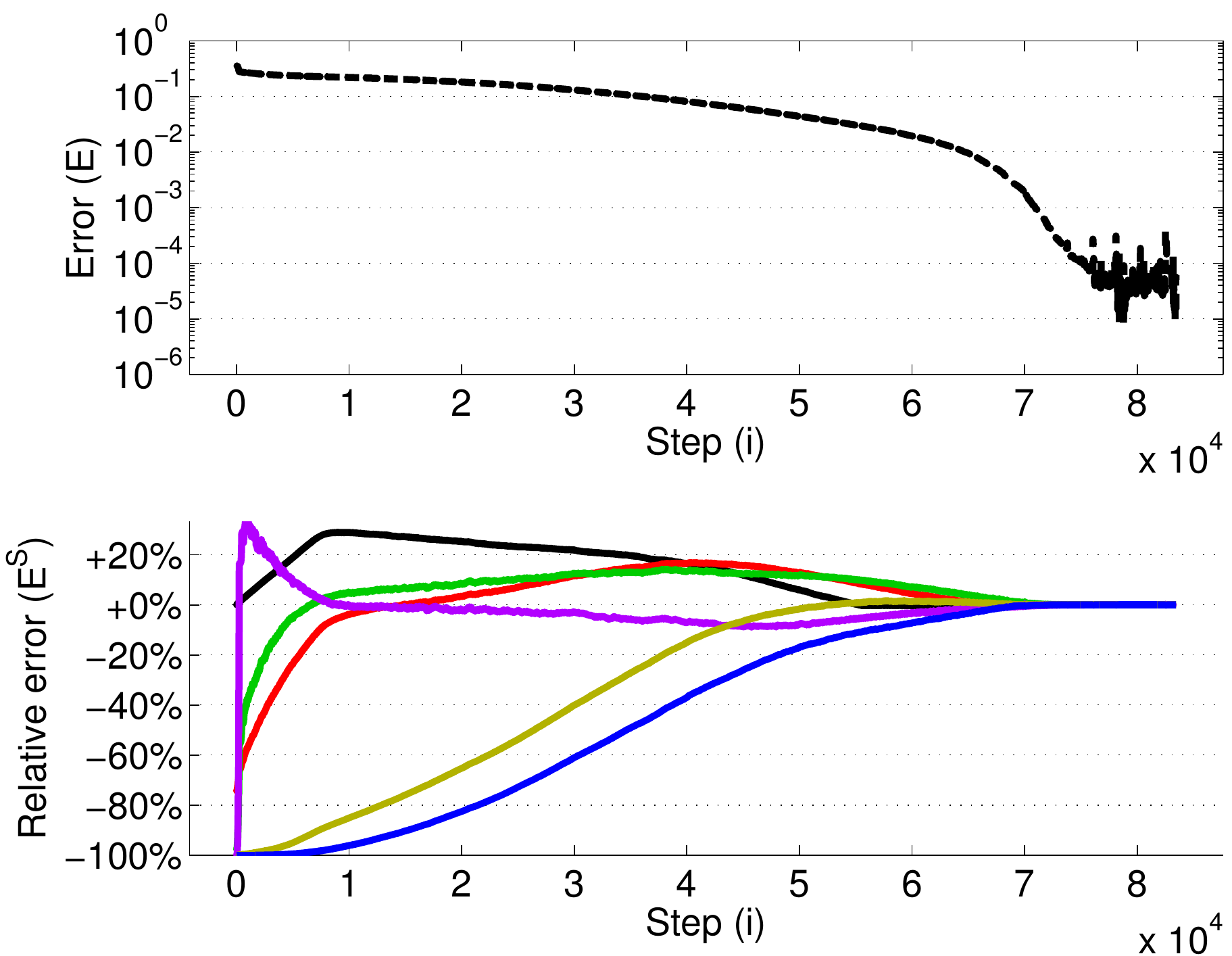}
  \includegraphics[width=0.6\columnwidth]{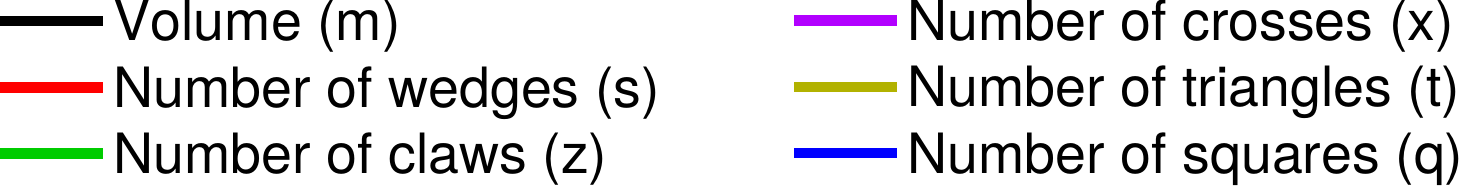}
  \caption{
    \label{fig:convergence}
    Convergence of the Guided Graph Generator for the \experimentNetwork
    network by \experimentNetworkCite. 
    Upper plot:  evolution of the error $E$.  Lower plot:  Evolution of
    the individual relative errors $E^S$. 
  }
\end{figure}

As to the convergence speed of the Guided Graph Generator,
we may interrupt the
algorithm at any timepoint to get a generated graph that is not optimal
in terms of our stopping criteria, but is faster to compute.  
Although the algorithm reduces the error $E$ over time, it is a priori
not clear
whether the algorithm converges at all, and
how changes for each individual statistic contribute to the
error. 
Thus, we show in Figure~\ref{fig:convergence} the
overall error 
$E$ (top plot) and the relative error $E^S$ for each statistic $S$
(lower plot) for the \experimentNetwork network.  
We observe phases in which each statistic is optimized, while in
some phases some statistics do not improve, and sometimes even regress.   
In fact, the number of edges, which is initially correct due to the
algorithm starting with an Erdős--Rényi network, increases at first to
allow the other count statistics to be corrected.  Once the other
statistics approach their intended values, the number of edges decreases
back to its correct value. The three $k$-star counts (wedges, claws and
crosses) increase fast at first, and also surpass their target values
only to come back later, allowing the triangle and square count to be
adjusted over time.
We note that the number of $k$-stars increases faster for higher $k$,
which we explain by the fact that adding edges to a high-degree node
makes them grow faster, which indicates that in a first phase, the
degree distribution is adjusted by the algorithm, while the clustering
statistics (the number of triangles and squares), reach their intended
values much more slowly in a second phase. 
Figure~\ref{fig:convergence} also shows the influence of the parameter
$\epsilon$ on the runtime of the proposed algorithm, as the number of steps $i$
is linear in the runtime of the algorithm.  The experiment shows that
the algorithm has runtime sublinear in $\epsilon^{-1}$. 

\begin{figure}
  \centering
  \includegraphics[width=0.8\columnwidth]{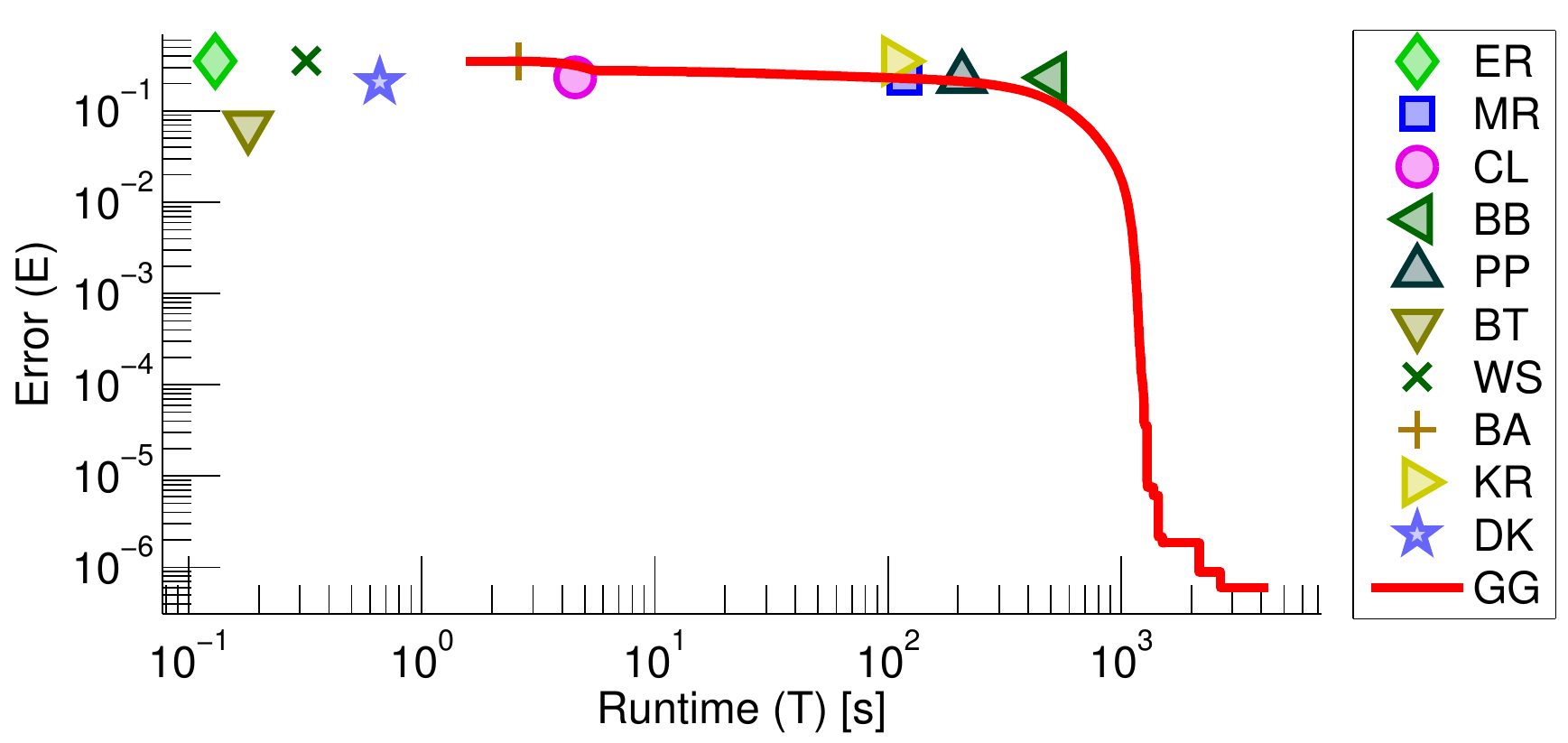}
  \caption{
    \label{fig:convergence2}
    The error $E$ in function of runtime for the \experimentNetwork
    network.  
    The proposed Guided Graph Generator is shown as a curve, reflecting the error of the
    current solution at different timepoints.  Other algorithms are
    shown as points. 
    Note that the Guided Graph Generator has by far the smallest error out of all
    algorithms; it is however not the fastest. 
  }
\end{figure}

In terms of convergence speed, 
we may also compare the Guided Graph Generator to other algorithms directly in two
ways:  (1) Does our algorithm generate a better graph when given a
runtime equal to that of another algorithm?  (2)~Is our algorithm faster
than other algorithms when generating graphs of equal quality?  Both
questions can be answered by inspecting the error $E$ in function of
runtime of the proposed algorithm, and comparing it to the runtime and error of
other algorithms.  This experiment is shown for the
\experimentNetwork network in 
Figure~\ref{fig:convergence2}. 
First of all, we can observe that the Guided Graph Generator is slower and more
precise than the other tested algorithms.  This is true not only for the
one shown network, but for all \experimentNetworkCount networks that we tested. 
Comparing other algorithms with ours at their runtime, or at their
precision shows that only BTER and the \textit{dK} model are consistently faster
at equal precision and more precise at the same runtime.  Note however
that these models use a $O(n)$- and $O(n^2)$-dimensional parameter space.

\section{Limitations of the Approach}
\label{sec:limitations}
Beyond the numerical results presented in the preceding section, we 
address here several important issues related to the proposed Guided
Graph Generator algorithm and to
graph generators based on network statistics in general. 

\subsection{Precision vs Underlying Models}
A common criterion for evaluating graph generators is the motivation for
the specific underlying graph generation rules.  For instance, the
Barabási--Albert model is very well justified by the principle of
preferential attachment, which is (at least conceptually) a phenomenon
believed to exist in the growth of real networks, and thus gives it a
justification:  Even if the Barabási--Albert model does not produce the
most realistic graphs (our experiments show that several other
algorithms beat it at that task), it nevertheless allows one to
understand \emph{why} real-world networks have the properties that they
have by proposing a mechanism that leads to them.  This is specifically
not the case for the Guided Graph Generator, which takes the
properties of an existing graph as input and reproduces these in a
synthetic graph.  In fact, we must stress that the iteration steps
performed by the algorithm do not correspond to an actual
graph evolution phenomenon, but purely to an optimization procedure.
Therefore, we come to the conclusion that the 
algorithm is not a graph \emph{model}, but instead a graph
\emph{generator}, a task at which it outperforms other algorithms. 

\subsection{Comparison of Statistics}
The Guided Graph Generator proposed in this paper indeed generates synthetic graphs with a high
precision~--  
both in terms of the properties it optimizes, as well as for other
properties such as the average distance, the assortativity and the
spectral norm.  The properties that it does not reproduce well are those
for which no algorithm yet exists that reproduces them:  the algebraic
connectivity $a$, the spectral distribution of the normalized adjacency
matrix $\mathbf N$, and the distance distribution.  
However, the proposed algorithm does not perform as well as Kronecker 
graphs in terms of the diameter. 
Also, specific other algorithms
that optimize for specific properties are better at reproducing these
properties:  the degree distribution for the algorithms of Molloy--Reed
and Bansal et al., the clustering coefficient for BTER, and the
assortativity for the \textit{dK} model.  

In our experiments, we also made notable observations on existing
network models.  For instance both the Kronecker and the \textit{dK} models produce
very unrealistic clustering coefficient distributions, even though at
least the Kronecker model includes clustering by way of its base matrix. 

\subsection{Precision vs Speed}
In terms of precision vs speed the Guided Graph Generator is very clearly placed on
the \emph{precision} side.  In fact, even algorithms known to be slow
such as Kronecker fitting are faster than the proposed algorithm in
many cases.  We note that the algorithm could be made faster simply by
reimplementing it in C or C++; we did not do this as our main concern was
precision~-- the fact that runtime was comparable between algorithms is
enough to make the algorithm tractable for many practical applications. 
Another avenue for improving the runtime of the algorithm is by
parallelization. 
First, the inner loop of the algorithm over all statistics $S\in
\mathcal S$ can be executed in parallel and second, 
each updating step uses only local information, and thus also lends
itself to parallelization. 

\subsection{Conflicting or Impossible Inputs}
In our examples and experiments, we have taken the subgraph counts
used as input to the Guided Graph Generator from actual given
real-world graphs.  In fact, Algorithm~\ref{alg:gg} as described
in this paper takes an input graph $G_0$ from which subgraph
counts are taken.  This makes sure that the target subgraph counts
optimized by the algorithm are realizable.  However, nothing prevents
us from using numbers as input which are not realizable.  For
instance, constructing a simple graph with 10 edges and 50 triangles is
an impossible task -- a graph with 10 edges can have at most 10 triangles;
this is realized by the complete graph on five nodes.  Faced with
such inputs, the Guided Graph Generator generates graphs that are
degenerate:  They don't come near the requested subgraph count values,
but instead are extremal, i.e., they often contain large cliques,
which are the most efficient way to create a large number of certain
subgraphs.  This can lead to certain interesting subcases:  When a
very large number of squares is requested at the same time than a very
small number of triangles, the algorithm converges to a complete
bipartite graph (i.e., a biclique). 
Note however that in general, it is very difficult to
determine whether a given combination of subgraph count is
realizable.  As an example, the general problem to determine whether
there exists a graph with a given number of nodes, edges and triangles
does not have a more efficient known solution that enumerating all
graphs with the requested properties. 

\subsection{Statistical Properties of the Guided Graph Generator}
For graph generation algorithms, it is a useful property to derive the
distribution of statistical values of the generated graphs when the
algorithm is run multiple times, or when the algorithm is used as a
sampling algorithm, i.e., subsequent graphs it generates are used as
output.  The proposed algorithm however does not try to generate a set
of graphs whose properties follow given distributions, but instead is
meant to find a single graph whose properties are as near as possible to
given target values.  As a result, the distribution of graph statistics,
if the algorithm were to be run for a large number of iterations, would
not converge towards any meaningful distribution, and thus cannot be
characterized as a Markov chain.  In fact, due to the greedy-like nature
of the algorithm, the distribution of statistics of generated graphs
would be distributed (in a most likely non-normal way) around the target
values.  Thus, the Guided Graph Generator cannot be used to sample
multiple graphs with a given distribution of statistics, but instead can
only be used to find a graph whose properties are as near as possible to
the target values. Note that an algorithm that generates a series of
graphs with a predictable distribution of properties will necessarily
produce graphs with a larger average error than an algorithm that
outputs only a single graph with minimal error, but cannot be used for
sampling.  The distribution of graph statistics which are not optimized
explicitly by the proposed algorithm thus cannot be derived, and thus
the experiments of Figure~\ref{fig:precision3} (bottom) are needed to
ascertain that the non-optimized statistics, too, have realistic values.

A further note can be made to compare the Guided Graph Generator with
approaches that use Gibbs sampling to generate graphs from an
exponential random graph model.  Characterizing the distribution of
statistics generated by a given iterative graph model is possible in
principle.  For instance, exponential random graph models (ERGMs) can be
generated by Gibbs sampling.  In an ERGM, the probability of each
individual graph $G$ is proportional to $\exp\{a_1 x_1(G) + a_2 x_2(G) +
\ldots \}$, where the $x_i(G)$ are the used graphs statistics, and the
values $\{a_i\}$ are the parameters of the model.  Thus, Gibbs sampling
can be implemented by asking, at each step of the iteration, whether an
edge should be flipped, and basing the decision on the ratio of
probabilities between the graph before and after the putative flip,
leading to an expression involving only the difference in the different
graph statistics.  While this iterative algorithm will generate graphs
from the desired exponential random graph model, it does not allow to
easily generate graphs that are similar to a given input graph: In order
to execute this algorithm, the parameters $a_i$ must be known, and in
order to determine these parameters, very computationally expensive
Monte-Carlo Markov chain estimation is necessary, as the only
relationship between the $a_i$ and the statistics of the target graph is
that they are related by a monotonously growing function \citep{b818}.
Thus, the Guided Graph Generator can be understood as bypassing the
issue of computing the parameter values $a_i$, and instead opting to use
the values of the original graph's statistics as parameters.  The 
price for this 
simplification however is that it is not anymore possible to
characterize the distribution of the generated graphs, but only to
measure empirically that their convergence is good enough for a given
practical application.

\section{Conclusion}
\label{sec:conclusion}
To summarize this article, we have presented an evaluation of common
graph generation algorithms in terms of numerical properties of the
graphs they generate, and shown that these common algorithms do not
perform ideally at that task, leading us to propose a novel algorithm
for it.  The proposed algorithm, the Guided Graph Generator, was shown
to beat previous algorithms at the task of generating networks with
specific values of network statistics, with the exception of graph
properties that are specifically optimized for by specific algorithms.
In particular, the proposed algorithm is able to generate graphs with a
given number of squares, and thus with a given bipartivity measure,
better than all other tested graph generators. 
On the other hand, we
were not able to beat Leskovec et al.'s Kronecker model in terms of the
generated graph's diameter.  Despite these 
positive results, the proposed algorithm is only a pure graph generator,
and must be carefully distinguished from the related but distinct
concept of graph models: By construction, the proposed algorithm does
not explain \emph{why} real-world networks have the properties that they
have.

In principle, the Guided Graph Generator algorithm as described in
Algorithm~\ref{alg:gg} can be applied to any numerical graph statistic,
making it possible to also optimize directly for instance for the
algebraic connectivity, diameter or assortativity.  In practice, doing
so is highly expensive, as new values of the statistics must be computed
for each node at each step.  In experiments, we were barely able to
generate such graphs based on the smallest network in our collection,
the Zachary karate club graph with 34 nodes.  Thus, this variant is
prohibited as long as no fast updating algorithms are available for
these statistics.  The same is true to a lesser extent for other count
statistics: While the expressions for the update of the number of crosses and squares
is tractable, higher star, cycle and clique counts require much more
complex expressions.  In particular, the ability to compute the count of
3-paths in an efficient way may make it possible to generate graphs with
more accurate values of the degree assortativity.

For the algebraic connectivity, the eigenvalues of the normalized
adjacency matrix $\mathbf N$, and the distance distribution, none of the
tested models reproduce realistic values, and it is an open problem to
formulate a graph model that fits each of them. 

\subsection*{Additional Information}

\noindent \textbf{Availability of Data and Material.}
The datasets used in the experiments are all part of the KONECT
project.\footnote{\url{http://konect.cc/}}  The names of the
dataset used are given in Table~\ref{tab:datasets}.  All datasets are
either available for download via the KONECT project, or available on
request from the authors, in case where a public distribution is not
allowed. 

\noindent \textbf{Funding.}
This research was partly funded by the EPSRC (EP/P004024 and Cambridge
University GCRF), and by the European Regional Development Fund
(ERDF/FEDER -- IDEES). 

\noindent \textbf{Authors' Contributions.}
The method was developed by J.\ Kunegis, J.\ Sun, and E.\ Yoneki. 
The experiments were implemented and executed by J.\ Kunegis.
The manuscript was written by J.\ Kunegis, J.\ Sun, and E.\ Yoneki. 

\noindent \textbf{Acknowledgments.}
We thank Valentin Dalibard and Christoph Schaefer for helpful comments on our
experiments, as well as for helping with executing third-party graph generator
implementations.  

\let\oldbibliography\thebibliography
\renewcommand{\thebibliography}[1]{%
  \oldbibliography{#1}%
  \setlength{\itemsep}{+0.16pt}%
}
\bibliographystyle{agsm}
\bibliography{ref,guided-graph-generation,konect,kunegis}

\end{document}